\documentclass[prb,aps,twocolumn,showpacs,superscriptaddress,amsmath,amssymb]{revtex4}
\usepackage{graphicx}
\usepackage{psfrag}
\usepackage{epsfig}
\usepackage{color}
\usepackage{amsmath}
\usepackage{subfigure}
\definecolor{dred}{rgb}{0.7,0.0,0.0}
\usepackage{dcolumn}
\usepackage{bm}

\begin{document}

\title {Microscopic origin of spin-orbital separation in Sr$_2$CuO$_3$}

 \author{Krzysztof Wohlfeld}
 \affiliation{Stanford Institute for Materials and Energy Sciences, SLAC National Laboratory and Stanford University, 2575 Sand Hill Road, Menlo Park, CA 94025, USA}
 \affiliation{Institute for Theoretical Solid State Physics, IFW Dresden, D-01069 Dresden, Germany}
 \author{Satoshi Nishimoto}
 \affiliation{Institute for Theoretical Solid State Physics, IFW Dresden, D-01069 Dresden, Germany}
 \author{Maurits W. Haverkort}
 \affiliation{Max-Planck-Institut f\"ur Festk\"orperforschung, D-70569 Stuttgart, Germany}
 \affiliation{Department of Physics and Astronomy, University of British Columbia, Vancouver, Canada V6T-1Z1}
\affiliation{Max Planck Institute for Chemical Physics of Solids,  D-01187 Dresden, Germany}
 \author{Jeroen van den Brink}
 \affiliation{Institute for Theoretical Solid State Physics, IFW Dresden, D-01069 Dresden, Germany}
 \affiliation{Department of Physics, TU Dresden, D-01062 Dresden, Germany}

\date{\today}

\begin{abstract}
Recently performed resonant inelastic x-ray scattering experiment (RIXS) at the copper $L_3$ edge in the quasi-1D Mott insulator Sr$_2$CuO$_3$ 
has revealed a significant dispersion of a single orbital excitation (orbiton). This large and unexpected
orbiton dispersion has been explained using the concept of spin-orbital fractionalization in which the orbiton, which is intrinsically
coupled to the spinon in this material, liberates itself from the spinon due to the strictly 1D nature of its motion.
Here we investigate this mechanism in detail by: (i) deriving the microscopic spin-orbital superexchange model 
from the charge transfer model for the CuO$_3$ chains in Sr$_2$CuO$_3$, (ii) mapping the orbiton motion in 
the obtained spin-orbital model into a problem of a single hole moving in an effective half-filled antiferromagnetic 
chain $t$-$J$ model, and (iii) solving the latter model using the exact diagonalization and obtaining the orbiton spectral function. Finally, the RIXS cross section is calculated based on the obtained orbiton spectral function and compared with the RIXS experiment.
\end{abstract}

\pacs{75.25.Dk, 75.30.Ds, 71.10.Fd, 78.70.Ck}

\maketitle

\section{Introduction}\label{sec:00}
{\it Long and difficult `search' for orbitons.---}
A relatively well-understood problem in strongly correlated electrons systems concerns
the propagation of collective magnetic (spin) excitations in Mott insulators
such as e.g. 3D LaMnO$_3$, 2D La$_2$CuO$_4$, ladder SrC$_2$O$_3$, and 1D Sr$_2$CuO$_3$~\cite{Imada1998}. The theoretically calculated
dispersion of such magnetic excitations (magnons in 2D or 3D, triplons in the ladder, or spinons in 1D)
agrees very well with the one measured using the 
inelastic neutron scattering~\cite{Coldea2001, Notbohm2007, Zaliznyak2004} or 
the resonant inelastic x-ray scattering (RIXS)~\cite{Ament2009, Haverkort2010, Braicovich2009a, Braicovich2009b}.
The origin of this fact is the relative simplicity of the spin-spin interactions, which are usually modeled
using the Heisenberg-type spin Hamiltonians~\cite{Imada1998, Coldea2001}. The excitation spectrum of such Hamiltonians can then be obtained 
using e.g. the linear spin wave approximation in 2D/3D~\cite{Auerbach1994} or the Bethe-Ansatz--based approaches in 1D~\cite{Auerbach1994, Giamarchi2004}.

This situation is very different when one considers propagation of the collective orbital
interactions -- the orbitons~\cite{Kugel1973} (coined as such in Ref. \onlinecite{Pen1997}). 
On the experimental side, this originates from the lack of experimental probe to measure orbiton dispersion~\cite{Ishihara2000, Forte2008, Marra2012}. 
Even if neutrons {\it do} couple to the orbital excitations~\cite{noteorbitalneutr} and can
in principle detect orbital waves~\cite{Ishihara2004}, this cannot be easily realized 
experimentally~\cite{Shamoto2005}. This is due to the {\it usually} low transfers of energy in the neutron scattering
experiments w.r.t. the energies needed to trigger the orbital excitations 
(for an exception see Ref. \onlinecite{Kim2011}).
Inelastic light scattering in the form of (optical) Raman scattering cannot transfer much momentum to orbital excitations leading
to controversial interpretations of the observed features~\cite{Saitoh2001, Grueninger2002, Ishihara2004}. Only recently it has been proposed~\cite{Ishihara2000, Forte2008} 
and then experimentally and theoretically established~\cite{Schlappa2012} that RIXS may be used to probe the orbitons' motion. Therefore, till last year, there were just three experimental 
indications of the existence of mobile orbitons:
(i) indirectly in the form of Davydov splittings~\cite{Macfarlane1971}  in Cr$_2$O$_3$, (ii)
more recently and also indirectly in a pump-probe experiment in the doped manganite~\cite{Polli2007}, and (iii) 
in the RIXS spectrum on titanates, where a very small (w.r.t. the experimental resolution)
dispersion was found~\cite{Ulrich2009}.

From the theoretical side  the situation is also complex. To understand the orbiton dispersion one has to take into account the interaction 
between orbitons and (i) the lattice (phonons) and (ii) the spin degrees of freedom.
Although the former has been investigated in several studies~\cite{Allen1999, vandenBrink2001, Schmidt2007} 
and for long `blamed' for causing a confinement of the orbiton motion~\cite{Allen1999, vandenBrink2001, Schmidt2007}, it turned out not to be of great importance in the here
discussed case of orbitons in Sr$_2$CuO$_3$~\cite{Schlappa2012}. Therefore, while still far from being understood, 
the interaction with the lattice will not be discussed in what follows. At the same time, however, the spin-orbital interaction~\cite{relspinorb}, 
which stems from the inherent entanglement of the spin and orbital degrees of freedom~\cite{Oles2006, Oles2012, Brzezicki2012, Brzezicki2013}
in the Kugel-Khomskii superexchange (and/or direct exchange)~\cite{Kugel1973, Kugel1982} models, which describe 
the propagation of spin or orbital excitations~\cite{Kugel1973}, has a profound impact on the orbiton motion~\cite{Feiner1997, Feiner1998, Ishihara1997, vandenBrink1998, Ishihara2000, Oles2000, Ishihara2004}.
Moreover, as already discussed in Refs.~\onlinecite{Khaliullin1997, Khaliullin2000, Kikoin2003, Wohlfeld2009, 
Herzog2011, Kim2012, Wen-Long2012, Kumar2012, Vieira2013}, and in direct relevance to the here discussed problem in Refs.~\onlinecite{Wohlfeld2011, Schlappa2012} (see also below), in order to correctly describe the collective orbital excitations, this interaction should not be treated on a mean-field level. This latter feature of the 
spin-orbital interaction severely complicates matter and is one of the main motivations for the study presented in this paper.

{\it Recent experimental and theoretical findings.---} This brief overview of the problems
with finding mobile orbitons, makes it clear that the recent experimental finding of the mobile orbiton
in Sr$_2$CuO$_3$~\cite{Schlappa2012} and its short theoretical description in Refs.~\onlinecite{Wohlfeld2011, Schlappa2012}, 
signifies a breakthrough in the study of orbital excitations.
We therefore briefly summarize these findings below.

The RIXS measurements performed at copper $L_3$ edge in Sr$_2$CuO$_3$~\cite{Schlappa2012} revealed two dispersive
orbital excitations (due to large crystal field splitting also called $dd$ excitations).
Firstly, the $d_{xz}$ orbital excitation, which (in the here used hole language) corresponds to a transfer
of a hole from the ground state $d_{x^2-y^2}$ orbital to the excited $d_{xz}$ excitation,
showed a sine-like dispersion. This dispersion was of the order of 200 meV, had a dominant $\pi$ period component,
and a large incoherent spectrum which lead to peculiar `oval'-like features in the RIXS spectrum, see Fig. 1 in Ref. ~\onlinecite{Schlappa2012}.
Secondly, also the $d_{xy}$ orbital excitation had a small dispersion with visible $\pi$ period component.
Finally, the other two orbital excitations (the $d_{yz}$ and the $d_{3z^2-r^2}$ orbital excitations) did not
show any significant dispersion.

While these experimental results are the first unambiguous observation
of an orbiton (cf. discussion above), they turned out to constitute a challenge from a theoretical perspective.
It was shown~\cite{Schlappa2012} that the above mentioned particular features of dispersion could only be explained
if the concept of spin-orbital separation was invoked and applied~\cite{Wohlfeld2011}.
In short, this concept suggests that: (i) the orbiton in Sr$_2$CuO$_3$ is so strongly
coupled to the spin excitations (spinons in this 1D case) that its coherent motion 
can only be explained if this coupling was explicitly taken into account, 
(ii) during its motion the orbiton can nevertheless `liberate' from the spinon.
This scenario can explain the reason why this orbiton
dispersion was not observed before: Since (on one hand) the spin-orbital separation phenomenon
is rather unique to 1D and (on the other hand) the experimental searches were constrained to mostly
2D or 3D compounds, the orbiton was finally only observed when the attention was turned into a purely 
1D system.

{\it Aim and plan of the paper.---}
In this paper we show how to apply the spin-orbital separation concept developed and discussed
in Refs.~\onlinecite{Wohlfeld2011, Schlappa2012} to the problem of the orbiton motion in Sr$_2$CuO$_3$.
We start from (Sec.~\ref{sec:0aa}) the proper charge transfer model for Sr$_2$CuO$_3$ supplemented by the terms which describe the dynamics of the excited orbitals. 
From this model we derive in Sec.~\ref{sec:0} the corresponding `Kugel-Khomskii' spin-orbital model which describes the spin and orbital dynamics in Sr$_2$CuO$_3$, and thus defines the Hamiltonian
that is used to calculate the orbiton spectral function. In Sec.~\ref{sec:1}, we calculate
the orbiton spectral function using the newly developed concept of spin-orbital separation~\cite{Wohlfeld2011}. Next, in Sec.~\ref{sec:2} we
establish the relation between the RIXS cross section and the orbiton spectral function calculated in Sec.~\ref{sec:1}
and compare the obtained RIXS spectra with those obtained in the experiment~\cite{Schlappa2012}.
Finally, in Sec.~\ref{sec:3} we discuss the possible other scenarios which might explain the experimental results presented in Ref.~\onlinecite{Schlappa2012} and end with the concluding remarks. 

The paper is supplemented by three appendices in which: (i) we discuss some details of the calculations performed in Sec.~\ref{sec:0a} (App. \ref{sec:applambda}) ,
(ii) we compare the results of Sec.~\ref{sec:1} and Sec.~\ref{sec:2} to those obtained using the linear orbital wave theory (App. \ref{sec:low}),
and (iii) we compare the results of Sec.~\ref{sec:2} with those obtained assuming all orbital excitations to be dispersionless (App.~\ref{sec:2a}).

\section{The charge transfer model}
\begin{figure}[t!]
   \includegraphics[width=0.5\textwidth]{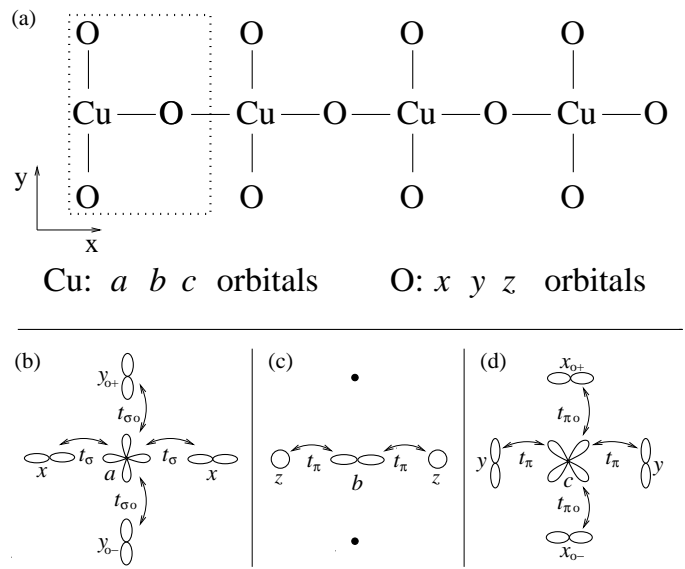}
\caption{The relevant atoms and orbitals that are taken
into account in model Eq. (\ref{eq:ct}): (a) orientation of
the CuO$_3$ chain with one Cu site and three O sites in the unit cell (dotted
line), (b) nearest neighbor O orbitals which hybridize with the Cu $a$ orbital 
following Eq. (\ref{eq:ct1}), (c) nearest neighbor O orbitals which hybridize with the Cu 
$b$ orbital following Eq. (\ref{eq:ct1}) -- note that there is no hybridization between
O orbitals that lie above / below the Cu-O-Cu-O-... chain, 
and (d) nearest neighbor O orbitals which hybridize with the Cu $c$ orbital 
following Eq. (\ref{eq:ct1}).
}
\label{fig:unitcell}
\end{figure}
\label{sec:0aa}

\subsection{Hamiltonian}
As noted in Sec. \ref{sec:00} the purpose of the present study 
is to describe the propagation of orbital excitation
in the quasi-1D cuprate Sr$_2$CuO$_3$~\cite{Ami1995, Motoyama1996, Kojima1997}.
Therefore, as our starting point we take the following 
multiband charge transfer Hamiltonian (which is an extended version of the charge transfer
model discussed in Ref. \onlinecite{Neudert2000}): 
\begin{equation}
\mathcal{{H}}= \mathcal{{H}}_0 + \mathcal{{H}}_1 + \mathcal{{H}}_2 + \mathcal{{H}}_3\,, 
\label{eq:ct} \\
\end{equation}
with $\mathcal{H}_0$ the tight binding Hamiltonian written in a basis consisting of five $3d$ orbitals
per Cu site and three $2p$ orbitals per O site. The many body interactions are included in $\mathcal{H}_1$, $\mathcal{H}_2$, and $\mathcal{H}_3$ 
(on-site Coulomb interaction on Cu atoms, on-site Coulomb interaction on O atoms, and nearest neighbor Coulomb interaction between electrons on Cu and O site).
In second quantized form and using the hopping parameters as indicated in Fig. \ref{fig:unitcell}
we obtain: \\
-- for the tight binding Hamiltonian $\mathcal{H}_0$
\begin{align}
\label{eq:ct1}
\mathcal{{H}}_0\!=
\!-t_{\sigma} \sum_{i, \sigma} &\!\Big( f^\dag_{ia \sigma} f_{i x \sigma}^{} -f^\dag_{i+1, a \sigma}f_{i x \sigma}^{}  +\mbox{H.c.} \Big) \nonumber \\
-t_{\sigma o} \sum_{i, \sigma} &\!\Big( f^\dag_{i a \sigma} f_{i yo+ \sigma}^{} -f^\dag_{i a \sigma}f_{i yo- \sigma}^{}  +\mbox{H.c.} \Big)
\nonumber \\
-t_{\pi} \sum_{i, \sigma} &\!\Big( f^\dag_{i c \sigma} f_{i y \sigma}^{} -f^\dag_{i+1, c \sigma}f_{i y  \sigma}^{}+ f^\dag_{i b \sigma} f_{i z \sigma}^{} \nonumber \\
&-f^\dag_{i+1, b 
\sigma}f_{i z  \sigma}^{} +\mbox{H.c.} \Big) \nonumber \\
-t_{\pi o} \sum_{i, \sigma} &\!\Big( f^\dag_{i c \sigma} f_{i xo+ \sigma}^{} -f^\dag_{i c \sigma}f_{i xo-  \sigma}^{} +\mbox{H.c.} \Big) \nonumber \\
+ \Delta_x \sum_{i} & (n_{ix} - n_{ia})+ \Delta_y \sum_{i} (n_{iy} - n_{ic}) \nonumber \\
 +  \Delta_z \sum_{i} & (n_{iz} - n_{ib}) 
 + \Delta_{yo} \sum_{i} (n_{iyo+} - n_{ia}) \nonumber \\
+ \Delta_{xo} \sum_{i} & (n_{iyo-} - n_{ia})+ \Delta_{xo} \sum_{i} (n_{ixo+} - n_{ib}) \nonumber \\
+ \Delta_{xo} \sum_{i} & (n_{ixo-} - n_{ib})   \nonumber \\
+ {\varepsilon}_a \sum_{i} &n_{i a } 
+ {\varepsilon}_b \sum_{i}  n_{i b } + {\varepsilon}_c \sum_{i} n_{i c },
\end{align} 
-- for the on-site Coulomb repulsion on copper sites term
\begin{align}
\mathcal{{H}}_1\!=\!&\sum_{i, \sigma, \alpha<\beta}(U-J^{\alpha\beta}_H)
n_{i \alpha \sigma}n_{i \beta {\bar\sigma}}
+\sum_{i, \sigma, \alpha<\beta}(U-2J^{\alpha\beta}_H) 
n_{i \alpha \sigma}n_{i \beta \sigma} \nonumber \\
&+U\sum_{i, \alpha}
n_{i \alpha \uparrow}n_{i \alpha \downarrow}
-\sum_{i, \sigma, \alpha<\beta} J^{\alpha\beta}_H f^\dag_{i \alpha \sigma} f_{i \alpha {\bar\sigma}}^{}
f^\dag_{i \beta {\bar\sigma}} f_{i \beta {\sigma}}^{} \nonumber \\
&+\sum_{i, \alpha<\beta} J^{\alpha\beta}_H f^\dag_{i \alpha \uparrow} f^\dag_{i \alpha \downarrow}
f_{i \beta \downarrow} f_{i \beta \uparrow},
\end{align}
-- for the on-site Coulomb repulsion on oxygen sites the Hamiltonian
\begin{align}
\mathcal{{H}}_2\!=\!&\sum_{i, \sigma, \mu<\nu}(U_p\!-\!J^{\mu\nu}_H)
n_{i \mu \sigma}n_{i \nu {\bar\sigma}} 
\!+\!\sum_{i, \sigma, \mu<\nu}\!(U_p\!-\!2J^{\mu\nu}_H)
n_{i \mu \sigma}n_{i \nu \sigma} \nonumber \\
&+U_p \sum_{i, \mu}
n_{i \mu \uparrow}n_{i \mu \downarrow}
-\sum_{i, \sigma, \mu<\nu} J^{\mu\nu}_H f^\dag_{i \mu \sigma} f_{i \mu {\bar\sigma}}^{}
f^\dag_{i \nu {\bar\sigma}} f_{i \nu {\sigma}}^{} \nonumber \\
&+\sum_{i, \mu<\nu} J^{\mu\nu}_H f^\dag_{i \mu \uparrow} f^\dag_{i \mu \downarrow}
f_{i \nu \downarrow} f_{i \nu \uparrow},
\end{align}
-- for the nearest neighbor Coulomb repulsion
\begin{eqnarray}
\mathcal{{H}}_3&\!=\!&V_{dp} \sum_{i, \mu \alpha} 
n_{i \alpha } ( n_{i \mu } + n_{i \mu + } +  n_{i \mu - } + n_{i+1 \mu }).
\end{eqnarray}
Here one needs to consider that
\begin{itemize}
\item the CuO$_3$ chain is oriented along the $x$ axis (Cu-Cu distance is set to 1), 
cf. Fig. \ref{fig:unitcell}(a); for simplicity hole notation is used;
\item the charge transfer model unit cell [cf. Fig. \ref{fig:unitcell}(a)] includes:
(i) one copper atom with three $3d$ orbitals:
$3d_{x^2-y^2}\equiv a$, $3d_{zx}\equiv b$, and
$3d_{xy}\equiv c$,
(ii) one oxygen atom within the Cu-O-Cu-O-... chain with three $2p$ orbitals:
$2p_x\equiv x$, $2p_y \equiv y$, $2p_z \equiv z$,
(iii) two equivalent oxygen atoms outside the Cu-O-Cu-O-... chain with two $2p$ orbitals:
above this chain -- $2p_x\equiv xo+$, $2p_y \equiv yo+$
and below this chain -- $2p_x\equiv xo-$, $2p_y \equiv yo-$;
\item the copper orbital indices are $\alpha, \beta \in \{a, b, c\}$,
the chain oxygen orbital indices are $\mu, \nu \in \{x, y ,z \}$, and the
spin index $\sigma \in \{ \uparrow, \downarrow \}$ ($\bar{\sigma}=-\sigma$);
\item $f_{i \kappa \sigma}$ annihilates a hole at site $i$  
in orbital $\kappa$ with spin $\sigma$ while
density operators are $n_{i \kappa} = n_{i \kappa \uparrow} + n_{i \kappa \downarrow}$
with $n_{i \kappa \sigma} = f^\dag_{i \kappa \sigma} f_{i \kappa \sigma}$;
\item the structure of the dominant hopping elements 
follows the Slater-Koster scheme (and was verified by our LDA calculations) and 
is depicted in Fig. \ref{fig:unitcell}(b)-(d); the rather large hopping between oxygens 
$t_{pp'}$ (cf. Ref. \onlinecite{Neudert2000}) is neglected; although this may  
give rise to a significantly smaller hole occupation on copper sites (and consequently may reduce the 
{\it intensity} of the dispersive $dd$ excitations), in the approach presented 
below it will not contribute to the superexchange processes;
\item the charge transfer energy $\Delta_\mu$ is measured for the particular 2$p$ orbital
from the relevant 3$d$ orbital (in the here used hole notation, see also above), 
i.e. from that 3$d$ orbital which hybridizes with this particular 2$p$ orbital;
\item due to crystal field there are distinct on-site energies $\varepsilon_\alpha$ for each 3$d$ orbital;
\item the structure of the Coulomb interaction follows 
Refs. \onlinecite{Griffith1964, Oles2005} and up to two-orbital interaction
terms exactly reproduces the correct on-site Coulomb interaction; 
note that the Coulomb interaction on oxygens above and below the chain is not considered
because we are interested merely in the Mott insulating case 
with one hole per copper site and in the 
analysis that follows {\it this particular} Coulomb interaction plays only a minor role.
\end{itemize}

We note at this point that the above 
model does not contain the $d_{yz} \equiv d $ and $d_{3z^2-r^2} \equiv e$ orbitals.
This is because there will not be any sizable dispersion due 
to the very small superexchange processes for the $dd$ excitations involving these orbitals. 
Note further that: (i) the hopping from
the $d_{yz}$ orbital to the neighboring oxygen along the chain direction
$x$ is negligible, and (ii) the hopping from the $d_{3z^2-r^2}$ to the $p_x$ orbital
on the neighboring oxygen is particularly small in this compound 
(much smaller than $t_\pi$ according to our LDA calculations).
We will therefore include these orbitals only when calculating the RIXS cross section
in Sec.~\ref{sec:2}.

\subsection{Parameters}
\begin{table*}[h!t] \caption{Assumed values of the parameters used
in the paper, see text for further details.
All parameters except
$R$, $R^m_n$ and $r^m_n$ (which are dimensionless) are given in eV.
Tilde before the value of the parameter denotes the fact that this 
precise value is not used 
in the analysis.} \label{tab:1}
\begin{ruledtabular}
\begin{tabular}{c c| c c | c c}
\multicolumn{4}{c|}{{\bf Model parameters}}                          &\multicolumn{2}{c}{{\bf Spectral function / RIXS parameters}}\cr \hline
\multicolumn{2}{c|}{Charge transfer model (Sec. \ref{sec:0aa})} & \multicolumn{2}{c|}{Spin-orbital model (Sec. \ref{sec:0})}  &\multicolumn{2}{c}{ 
Local excitation energies (Sec. \ref{sec:1}, \ref{sec:2}, and App. \ref{sec:2a}) } \cr \cline{1-6}
$t_{\sigma}$           & 1.5                                &   $J_1$       & 0.088             & $E_b$                & 2.15 \cr 
$t_{\pi}$              & 0.83                               &   $J^b_2$   &  0.021                  & $E_c$                & 1.41 \cr
$t_{\sigma o}$         & 1.8                                &   $J^c_2$   &  0.010           & $E_d$ & 2.06 \cr 
$t_{\pi o}$            & 1.0                                &   $J^b_{12}$&  0.043 &  $E_e$              &2.44 \cr 
$\Delta_x$             & 3.0                                &   $J^c_{12}$&  0.030   & $E_{AF}$ & 0.33 \cr \cline{5-6}
$\Delta_y$             & 2.8       &   $R$       &     1.7                     & \multicolumn{2}{c}{Effective $t$--$J$ models (Sec. \ref{sec:1}) } \cr \cline{5-6}
$\Delta_z$             & 2.2       &   $R^b_1$   & 2.5   & $ t_b $                & 0.084  \cr 
$\Delta_{xo}$          & 2.5                                &   $R^c_1$   &  2.3    & $t_c $                & 0.051  \cr
$\Delta_{yo}$          & 3.5                                &   $R^b_2$   &  2.0 & $J   $              &0.24   \cr \cline{5-6}
$U$                    & 8.8                                &   $R^c_2$   & 1.9      &       \multicolumn{2}{c}{ Linear orbital wave approximation (App. \ref{sec:low})  }    \cr \cline{5-6}
$U_p$                  & 4.4                                &   $r^b_1$ & $ 1.7 $                 & $B$ &  2.48  \cr 
$J^b_H \equiv J^{ab}_H$                & 1.2       & $r^c_1$      & $ 1.3 $                    & $C$   &  1.74  \cr 
 $J^c_H \equiv J^{ac}_H$                & 0.69           &   $r^b_2$         & 1.2             & $J_b$  &  -0.019   \cr 
 $J^p_H \equiv J^{\mu\nu}_H$                & 0.83      &   $r^c_2$         & 1.1          &  $J_c$          &  -0.014   \cr
 $V_{dp}$               &   1.0                              &   $\bar{\varepsilon}_a$          &   0.0                     &  & \cr
 ${\varepsilon}_a$  &   0.0                              &   $\bar{\varepsilon}_b$          &   $\sim 2.2$           &                  &  \cr
 ${\varepsilon}_b$  &  $\sim 0.5$                    &   $\bar{\varepsilon}_c$      &   $\sim 2.0$       &                                   &               \cr
 ${\varepsilon}_c$  &  $\sim 0.5$                   &                                         &                                  &                               & \cr
\end{tabular}
\end{ruledtabular}
\end{table*}

In the model Hamiltonian Eq. (\ref{eq:ct}) a large number of parameters appear, which need
to be fixed in order to obtain quantitative results that can be compared to experiment,
cf. left column of Table~\ref{tab:1}.

In principle we used the basic set of the parameters that was proposed
in Ref.~\onlinecite{Neudert2000}. The only exception is the intersite Coulomb repulsion $V_{dp}$ which
is set to a somewhat smaller value of 1 eV than the exceptionally large one suggested in Ref.~\onlinecite{Neudert2000}.
Note however, that the smaller value is still generally accepted for the cuprates~\cite{Grant1992}.
Besides, this value leads to the spin superexchange parameter that
is equal to 0.24 eV (see below), which is the experimentally
observed value~\cite{Schlappa2012}. Finally, we used the following values for parameters 
not considered in Ref.~\onlinecite{Neudert2000}:

(i) The Hund's exchange $J^c_H$ and $J^b_H$ are calculated using Slater integrals from 
Ref. \onlinecite{Haverkort2005} while $J^p_H$ is taken from Ref. \onlinecite{Grant1992}.

(ii) $\varepsilon_b$ and $\varepsilon_c$ are estimated
to be 0.5 eV from the LDA calculations.

(iii)  While following Ref. \onlinecite{Neudert2000} $\Delta_x = 3.0$ eV
and $\Delta_{y0}=3.5$ eV, the values of the other charge transfer
parameters are not given in this reference and have to be obtained in another way. 
It seems reasonable to assume first that values of 
$\Delta_{xo}$,  $\Delta_y$, and $\Delta_z$ are roughly of the order
of $\Delta_x$. But, since the charge transfer parameters
are defined as equal to the difference in energy between the particular {\it hybridizing} 
$c$ or $b$ orbital and the particular 2$p$ orbital, they have to be lower than $\Delta_x$. 
Quantum chemical calculations suggest, however,
that the actual values of $\Delta_y$ and $\Delta_z$ might still be different: it occurs that the values of the 
on-site energies of the $b$ and $c$ orbital are not identical and that the
$b$ orbital has a higher energy than the $c$ orbital by ca. 0.7 eV, 
cf. Ref. \onlinecite{Schlappa2012}.
Altogether, this suggests the following values for these two parameters:
$\Delta_y = 2.8$ eV and $\Delta_z = 2.2$ eV.
We will show later that these values give the orbiton dispersion in reasonably good
agreement with the RIXS experiment \cite{Schlappa2012}.

(iv) $t_{\pi}$ and $t_{\pi o}$ are assumed to be roughly of the order of 55\%
of $t_{\sigma}$ and $t_{\sigma o}$ (respectively)\cite{Grant1992}.

Note that Table~\ref{tab:1} contains also a few other parameters which are
later introduced in this paper. While they mostly follow from
the charge transfer model parameters mentioned above, 
we will comment on their origin once they become relevant
in the following sections.

\section{Derivation of the spin-orbital model}
\label{sec:0}

Since the Coulomb repulsion $U$ and the charge
transfer energies $\Delta_\mu$ present in model Eq. (\ref{eq:ct})
are far larger than the hoppings $t_n$ ($t_n \ll U$ and $t_n \ll \Delta_\mu$ where $n = \sigma, \pi, \sigma o, \pi o$),
cf. Table~\ref{tab:1}, the ground state of $\mathcal{H}$ is a Mott insulator.
This is because, in the zeroth order approximation in the perturbation theory
in hopping $t_n$ and in the regime
of one hole per copper site, there is one hole localized in the $a$ orbital 
at each copper site $i$. Similarly, when a single orbital excitations is made, 
then in the zeroth order 
the hole will be localized on a single copper site in a particular
$b$ or $c$ orbital (because the charge transfer energy $\Delta_\mu$ is always positive). 

In the second and fourth order perturbation theory in $t_n$ (the terms obtained from $t_n$ and $t_n^3$ 
perturbation vanish~\cite{Zaanen1988}) the hole can 
delocalize which leads to a particular low energy Hamiltonian -- the spin-orbital Hamiltonian. 
This Hamiltonian has the following generic structure:
\begin{align} \label{eq:hspinorb}
\mathcal{\bar{H}} = \mathcal{\bar{H}}_0 + \mathcal{\bar{H}}_a + \mathcal{\bar{H}}_b + \mathcal{\bar{H}}_c.
\end{align}
It consists of two kinds of terms:
(i) $\mathcal{\bar{H}}_0$ which is a result of the second order perturbation theory in $t_n$, 
and (ii) $\mathcal{\bar{H}}_a + \mathcal{\bar{H}}_b + \mathcal{\bar{H}}_c$ terms which
follow from the fourth order perturbation theory in $t_n$ and can be called `superexchange' terms.
Note that the latter terms can be classified in two classes:
(i) $\mathcal{\bar{H}}_a$ -- the so-called `standard' or `spin' superexchange terms, which contribute 
when all holes are in the $a$ orbitals (i.e. no orbital excitations are present), and
(ii) $\mathcal{\bar{H}}_b + \mathcal{\bar{H}}_c$ -- the spin-orbital superexchange (`Kugel-Khomskii'--like) 
terms with one orbital excitation present on one site
of the bond (in $b$ or $c$ orbital) and no orbital excitation present on the other site
of the bond. In the following subsections we discuss these terms `step-by-step'.

\subsection{Renormalization of on-site energies: $\mathcal{\bar{H}}_0$}
\label{sec:0a} 

In the second order perturbation theory in $t_n$ the hole can delocalize to
the four neighboring oxygen sites surrounding the copper 
sites forming bonding and antibonding states. 
Although there are many important consequences
of such $t_n^2$ processes, 
let us now just explore one of them which actually 
turns out to be very important: the renormalization
of the on-site energies of the orbitals.
In Appendix \ref{sec:applambda} we discuss another, perhaps less important, consequence
of these processes: the renormalization of the hopping 
within the chain due to hybridization with oxygen orbitals above and below the chain
(these renormalization factors are called $\lambda_a$ and $\lambda_c$).

When the hole delocalizes into the bonding
and antibonding states formed by the $a$, $b$, or $c$ orbitals
with the four surrounding oxygen sites, the effective on-site energies
of the orbital levels are strongly renormalized with respect to the energy levels of
the pure $a$, $b$, or $c$ orbitals. Although the proper calculation of this phenomenon
can be done analytically by diagonalizing
a five level problem defined separately for each
of the copper $\alpha$ orbitals, we do not perform it here.
Instead we take the values obtained from the Ligand 
Field Theory Programme~\cite{Haverkort2012} based on the multiplet
ligand field theory using Wannier orbitals
on a CuO$_4$ cluster. 
It occurs that, for realistic values of parameters of model Eq. (\ref{eq:ct}) (see Table~\ref{tab:1}), 
the antibonding states are well-separated from the 
bonding states and we can safely neglect the
latter ones in the low energy limit that is of interest here. This leads
to the following term in our spin-orbital model:
\begin{equation} \label{eq:new}
\mathcal{\bar{H}}_0=\bar{{\varepsilon}}_a \sum_{i} \tilde{n}_{i  a } + \bar{{\varepsilon}}_b \sum_{i} \tilde{n}_{i  b } 
+ \bar{{\varepsilon}}_c \sum_{i} \tilde{n}_{i  c },
\end{equation}
where the values of the parameters $\bar{\varepsilon}_\alpha$ are
shown in Table~\ref{tab:1}. Here, we use the operators $n_{i \alpha}$ from
Eq.~(\ref{eq:ct}), although a rigorous treatment would require the use
of the operators actually creating the particular bonding states centered around 
a copper $\alpha$ orbital at site $i$. We discuss in Appendix \ref{sec:applambda}
why such simplification is to a large extent justified. Besides,
the tilde above the operators denotes the fact
that we prohibit double occupancies in this low energy Hamiltonian
due to the large on-site Hubbard $U$ and $U_p$.

It is convenient to define at this point the orbital pseudospin operators:
 ${\bf \sigma} =\frac12$ where
\begin{align} \label{eq:pseudospinb}
\sigma^z_i &= \frac12 (\tilde{n}_{ib}-\tilde{n}_{ia}), \nonumber \\
\sigma^+_i &= \tilde{f}^\dag_{b i} \tilde{f}_{a i}, \nonumber \\ 
\sigma^-_i &= \tilde{f}^\dag_{a i} \tilde{f}_{b i}.
\end{align}
and ${\bf \tau} =\frac12$ where
\begin{align} \label{eq:pseudospinc}
\tau^z_i &= \frac12 (\tilde{n}_{ic}-\tilde{n}_{ia}), \nonumber \\  
\tau^+_i &= \tilde{f}^\dag_{c i} \tilde{f}_{a i}, \nonumber \\
\tau^-_i &= \tilde{f}^\dag_{a i} \tilde{f}_{c i},
\end{align}
Here the tilde above the operators denotes the fact
that double occupancies are forbidden in this low energy Hamiltonian
due to large on-site Coulomb repulsion $U$ and $U_p$.
Setting $  \bar{{\varepsilon}}_a=0$ we can rewrite Eq. (\ref{eq:new}) as follows
\begin{equation}  \label{eq:new1}
\mathcal{\bar{H}}_0= \bar{{\varepsilon}}_b \sum_{i} \Big(\frac12+ \sigma^z_i \Big)
+ \bar{{\varepsilon}}_c \sum_{i} \Big(\frac12 + \tau^z_i \Big).
\end{equation}

\subsection{Spin superexchange: $\mathcal{\bar{H}}_a$}
\label{sec:0b}
\begin{figure*}[t!]
\centering
   \includegraphics[width=2.00\columnwidth]{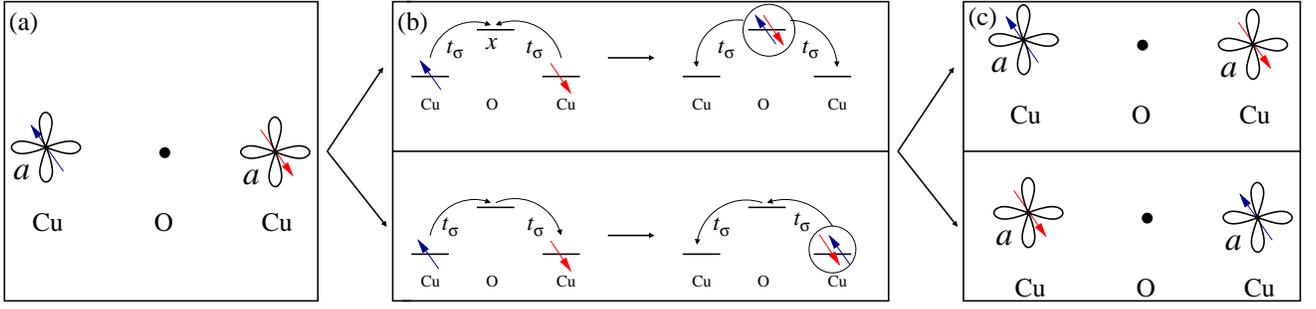}
\caption{(Color online) Schematic view of superexchange interactions when no orbital excitations
are present, i.e. between spins (arrows) of the holes in the $a$ orbitals:
the `initial' state [panel (a)]
can be brought into the so-called `virtual' state
of the superexchange process [circle with two spins (arrows) on the 
right side of panel (b)] by the virtual hopping [left side of panel (b)] 
which can `decay' [right side of panel (b)] via the virtual hopping
into the `final' state of the superexchange [panel (c)].
Panel (b) shows two kinds of `virtual' states with doubly occupied ions:
on the oxygen (copper) on the lower (upper) side of the panel
with the energy cost $\propto U$ ($\propto Up+2\Delta$), respectively. 
Panel (c) shows two possible low energy `final' configurations
without double occupancies: without spin flip (i.e. identical
to the initial state) and with spin flip.
Note that panel (b) shows relevant copper and oxygen orbitals in a schematic
way, i.e. depicted by horizontal bars, while panels (a) and (c)
explicitly show the relevant copper orbitals (oxygen orbitals are not
shown on these panels).
}
\label{fig:AF_SE}
\end{figure*}

Let us firstly study the superexchange interactions when only one type
of orbital is occupied along the superexchange bond.
In this case it is very straightforward to show that
model Eq. (\ref{eq:ct}) can be easily reduced to the low
energy Heisenberg model for spins $S=1/2$ using the perturbation
theory to fourth order in ${t}_{\sigma}$~\cite{Zaanen1988}, cf. 
Fig.~\ref{fig:AF_SE}:
\begin{align}
\label{eq:h1}
\mathcal{\bar{H}}_a&=J_1 (1+R) \sum_{i } \mathcal{P}_{i, i+1} \left({\bf S}_{i } \cdot {\bf S}_{i+1} -
\frac{1}{4} \right), 
\end{align}
where $\mathcal{P}_{i, i+1}$ denotes the fact that there are no orbital excitation 
present along the bond $\langle i, i+1 \rangle$ and is defined as
\begin{align}
\mathcal{P}_{i, i+1}= \Big( \frac12 + \tau^z_i \Big) \Big( \frac12 + \tau^z_{i+1} \Big) \Big( \frac12 + \sigma^z_i \Big) \Big( \frac12 + \sigma^z_{i+1} \Big).
\end{align}
Note that here the superexchange interactions involve not only the spin degree 
of freedom $S =\frac12$:
\begin{align} \label{eq:spin}
S^z_i &= \frac12 (\tilde{n}_{i \uparrow} - \tilde{n}_{i \downarrow}), \nonumber \\
S^+_i &= \tilde{f}^\dag_{i \uparrow} \tilde{f}_{\downarrow i}, \nonumber \\
S^-_i &= \tilde{f}^\dag_{i \downarrow} \tilde{f}_{\uparrow i}.
\end{align}
but also the orbital degree of freedom pseudospin operators defined in the previous subsection. 
Again the tilde above the operators denotes the fact
that double occupancies are forbidden in this low energy Hamiltonian
due to large on-site Coulomb repulsion $U$ and $U_p$.

The superexchange constant contains contributions due to charge excitations 
on copper sites ($\sim J_1$) and on the oxygen sites located in between
the copper sites ($\sim J_1 R$), where
\begin{equation} \label{eq:j1}
J_1=\left(\frac{2\bar{t}^2_{\sigma}}{\Delta_x+V_{dp}}\right)^2 \frac{1}{U},
\end{equation}
with $\bar{t}_{\sigma} = \lambda_a {t}_{\sigma}$ (see Appendix \ref{sec:applambda} for origin
of the factor $\lambda_a$) and
\begin{equation}
R=\frac{2U}{2\Delta_x+U_p}.
\end{equation}

Two remarks are in order here. Firstly, when no orbital excitations are present, 
the Hamiltonian Eq. (\ref{eq:h1}) is equal to the the well-known spin-only Heisenberg model.
This is in agreement with the `common wisdom' stating that the orbital degrees of freedom can be
easily integrated out in systems with only one orbital occupied in the ground state.
Secondly, in the above derivation we neglected intermediate states with ${}^1A_1$ or ${}^1E$
symmetry. In principle superexchange processes which involve these intermediate states should
also be taken into account. However, due to the crystal field splitting 
this would mean that the final states of the superexchange process would
contain high energy orbital excitations. Consequently 
these processes are suppressed. 

\begin{figure*}[t!]
\centering
   \includegraphics[width=2.00\columnwidth]{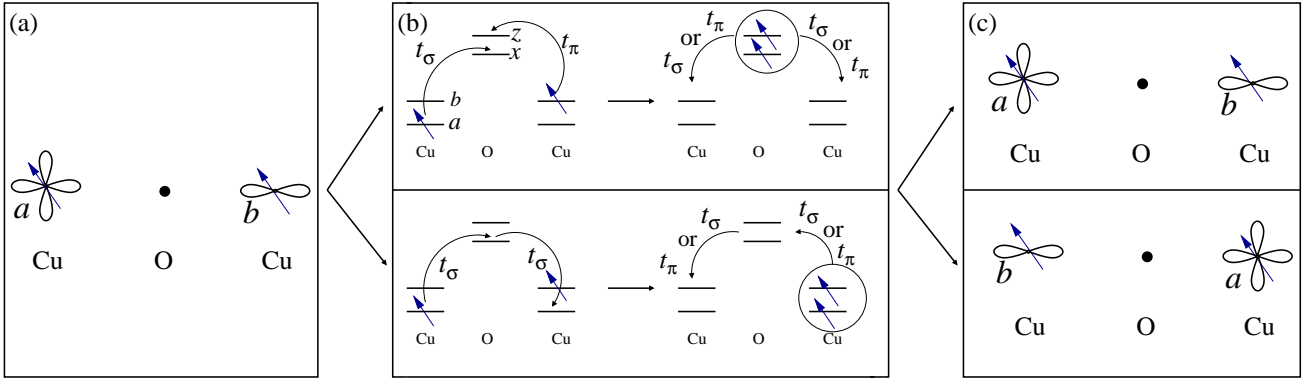}
\caption{(Color online) Schematic view of superexchange interactions with
one orbital excitation (here: into orbital $b$) on one site 
and no orbital excitations on the other site (hole in orbital $a$)
in the case that both spins (arrows) on the neighbouring orbitals 
are {\it parallel}: similarly as in Fig. \ref{fig:AF_SE} 
the `initial' state [panel (a)]
can be brought into the so-called `virtual' state
of the superexchange process [circle with two spins (arrows) on the 
right side of panel (b)] by the virtual hopping [left side of panel (b)] 
which can `decay' [right side of panel (b)] via the virtual hopping
into the `final' state of the superexchange [panel (c)].
Panel (b) shows two kinds of `virtual' states with doubly occupied ions --
on the oxygen and copper site, cf. Fig. \ref{fig:AF_SE}. 
Panel (c) shows two possible low energy `final' configurations
without double occupancies: without {\it orbital} flip (i.e. identical
to the initial state) and with {\it orbital} flip.
}
\label{fig:AOFM_SE}
\end{figure*}
%

\begin{figure*}[t!]
\centering
   \includegraphics[width=2.00\columnwidth]{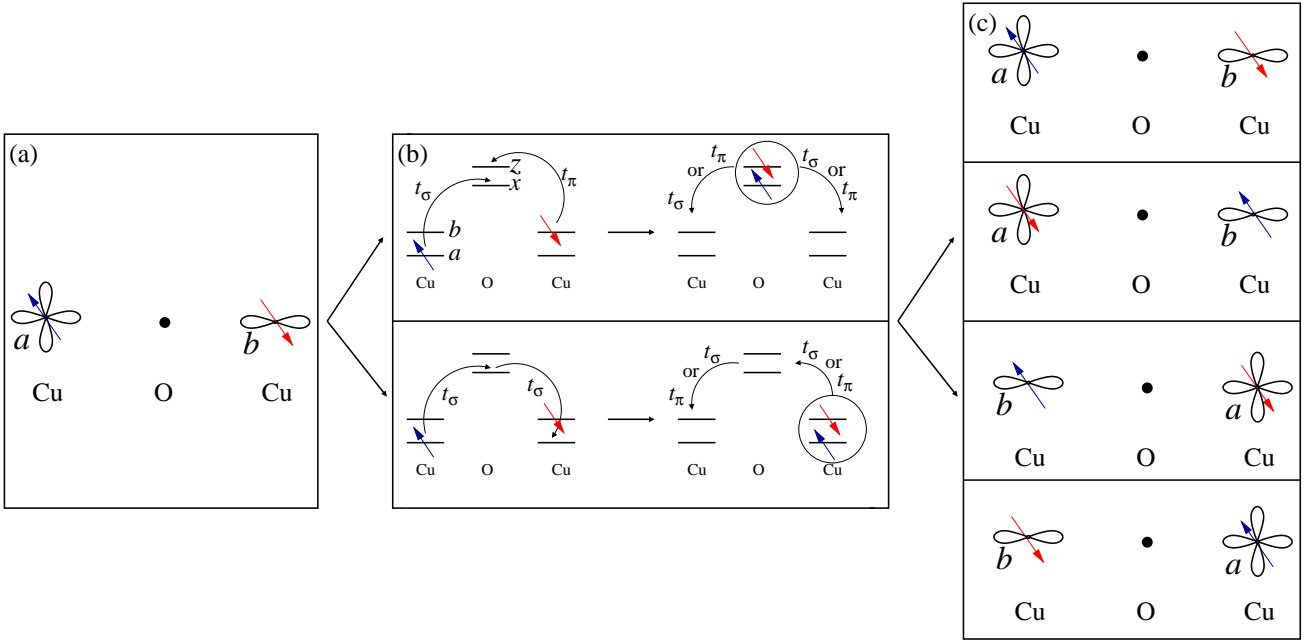}
\caption{(Color online) Schematic view of superexchange interactions with
one orbital excitation (here: into orbital $b$) on one site 
and no orbital excitations on the other site (hole in orbital $a$)
in the case that spins (arrows) on the neighboring orbitals 
are {\it antiparallel}: similarly as in Fig. \ref{fig:AF_SE} 
the `initial' state [panel (a)]
can be brought into the so-called `virtual' state
of the superexchange process [circle with two spins (arrows) on the 
right side of panel (b)] by the virtual hopping [left side of panel (b)] 
which can `decay' [right side of panel (b)] via the virtual hopping
into the `final' state of the superexchange [panel (c)].
Panel (b) shows two kinds of `virtual' states with doubly occupied ions --
on the oxygen and copper site, cf. Fig. \ref{fig:AF_SE}. 
Panel (c) shows {\it four} possible low energy `final' configurations
without double occupancies: with/without {\it orbital} flip 
and with/without {\it spin} flip.
}
\label{fig:AOAF_SE}
\end{figure*}

\subsection{Spin-orbital superexchange for $b$ orbital: $\mathcal{\bar{H}}_b$}
If along a bond there is one hole in the $b$ orbital 
(due to e.g. an orbital excitation created in RIXS) and another one in the $a$ orbital, then
using the perturbation theory to fourth order in $t_n$ we obtain (cf. Figs \ref{fig:AOFM_SE}-\ref{fig:AOAF_SE}):
\begin{align}
\label{eq:hb}
\mathcal{\bar{H}}_b= &\sum_{i } \left({\bf S}_{i } \cdot {\bf S}_{i+1} +\frac{3}{4}\right) 
\Big[\left(R^b_1 J^b_{12}+r^b_1 \frac{J_1+J^b_2}{2}\right) \nonumber \\
&\left(\sigma^z_i \sigma^z_{i+1} - \frac{1}{4} \right) 
+\frac{R^b_1+r^b_1}{2} J^b_{12} \left( \sigma^+_i \sigma^-_{i+1} + \sigma^-_i \sigma^+_{i+1} \right) \Big] \nonumber \\
+& \sum_{i } \left(\frac{1}{4}- {\bf S}_{i } \cdot {\bf S}_{i+1}\right) 
\Big[\left(R^b_2 J^b_{12}+r^b_2 \frac{J_1+J^b_2}{2}\right) \nonumber \\
&\left(\sigma^z_i \sigma^z_{i+1}-\frac{1}{4}\right) 
-\frac{R^b_2+r^b_2}{2} J^b_{12} \left( \sigma^+_i \sigma^-_{i+1} + \sigma^-_i \sigma^+_{i+1} \right) \Big],
\end{align}
where the superexchange constant $J_1$ is the one given by Eq. (\ref{eq:j1}) while
\begin{equation}
J^b_2=\left(\frac{2{t}^2_{\pi}}{\Delta_z+V_{dp}}\right)^2 \frac{1}{U},
\end{equation}
and
\begin{equation}
J^b_{12}=\frac{\left(2{t}_{\pi}\bar{t}_{\sigma}\right)^2}{(\Delta_z+V_{dp})(\Delta_x+V_{dp})} \frac{1}{U}.
\end{equation}

The complex structure of Hamiltonian (\ref{eq:hb}) is a consequence
of the fact that the proper derivation of such low energy
model has to include the superexchange processes with
four distinct intermediate states:

(i) The high spin state ${}^3T_1$ on copper sites (middle bottom panel of Fig. \ref{fig:AOFM_SE}) which involves $d^2 = b^1 a^1$ orbital configuration 
and  with energy in terms of Racah parameters
\cite{Griffith1964} $A-5B \equiv U- 3J^b_H$. This leads to $r^b_1 = \frac{1}{1-3J^b_H/U}$.

(ii) The low spin state ${}^1T_1$ on copper sites (middle bottom panel of Fig. \ref{fig:AOAF_SE}) which involves $d^2 = b^1 a^1$ orbital configuration 
and  with energy in terms of Racah parameters
\cite{Griffith1964} $A+B +2C \equiv U- J^b_H$. This leads to 
$r^b_2 = \frac{1}{1-J^b_H/U}$.

(iii) The high spin state ${}^3T_1$ on oxygen sites (middle top panel of Fig. \ref{fig:AOFM_SE}; 
note that due to the equivalence between the $t^2_{2g}$ and $p^2$ configuration \cite{Griffith1964} we can label the multiplet states on the $p$ shell by those known from the $t_{2g}$ sector) which involves $p^2 = x^1 z^1$ orbital configuration and with energy in terms of Racah parameters
\cite{Griffith1964} $A_o-5B_o \equiv U_p- 3J^p_H$ (where $o$ denotes the fact that the Racah parameters are for oxygen sites). 
This leads to $R^b_1 = \frac{2U}{\Delta_x+\Delta_z+U_p(1-3J^p_H/U_p)}$.

(iv) The low spin state ${}^1T_2$ on oxygen sites (middle top panel of Fig. \ref{fig:AOAF_SE}) which involves $p^2 = x^1 z^1$ orbital configuration 
and  with energy in terms of Racah parameters
\cite{Griffith1964} $A_o+B_o+2C_o \equiv U_p- J^p_H$ . This leads to $R^b_2 = \frac{2U}{\Delta_x+\Delta_z+U_p(1-J^p_H/U_p)}$.

\subsection{Spin-orbital superexchange for $c$ orbital: $\mathcal{\bar{H}}_c$}
If along a bond there is one hole in the $c$ orbital (due to e.g. an orbital
excitation created in RIXS) and another one in the $a$ orbital, then
using the perturbation theory to fourth order in $t_n$ we obtain (cf. Figs \ref{fig:AOFM_SE}-\ref{fig:AOAF_SE} showing an analogous situation in the case of 
the orbital superexchange between the $b$ and $a$ orbitals):
\begin{align}
\label{eq:hc}
\mathcal{\bar{H}}_c= &\sum_{i } \left({\bf S}_{i } \cdot {\bf S}_{i+1} +\frac{3}{4}\right) 
\Big[\left(R^c_1 J^c_{12}+r^c_1 \frac{J_1+J^c_2}{2}\right) \nonumber \\
& \left(\tau^z_i \tau^z_{i+1} - \frac{1}{4} \right) +\frac{R^c_1+r^c_1}{2} J^c_{12} \left( \tau^+_i \tau^-_{i+1} + \tau^-_i \tau^+_{i+1} \right) \Big] \nonumber \\
+& \sum_{i } \left(\frac{1}{4}- {\bf S}_{i } \cdot {\bf S}_{i+1}\right) \Big[\left(R^c_2 J^c_{12}+r^c_2 \frac{J_1+J^c_2}{2}\right) \nonumber \\
& \left(\tau^z_i \tau^z_{i+1}-\frac{1}{4}\right) -\frac{R^c_2+r^c_2}{2} J^c_{12} \left( \tau^+_i \tau^-_{i+1} + \tau^-_i \tau^+_{i+1} \right) \Big],
\end{align}
where the superexchange constants are $J_1$ [cf. Eq. (\ref{eq:j1})] and
\begin{equation}
J^c_2=\left(\frac{2\bar{t}^2_{\pi}}{\Delta_y+V_{dp}}\right)^2 \frac{1}{U},
\end{equation}
with $\bar{t}_{\pi} = \lambda_c {t}_{\pi}$ (see Appendix \ref{sec:applambda} for origin
of the factor $\lambda_c$) and
\begin{equation}
J^c_{12}=\frac{\left(2\bar{t}_{\pi}\bar{t}_{\sigma}\right)^2}{(\Delta_y+V_{dp})(\Delta_x+V_{dp})} \frac{1}{U}.
\end{equation}

Similarly to the $b$ orbital case discussed above, the complex structure of Hamiltonian (\ref{eq:hc}) is a consequence of the fact that the proper derivation of such low energy
model has to include the superexchange processes with
four distinct intermediate states:

(i) The high spin state ${}^3T_1$ on copper sites which involves $d^2 = c^1 a^1$ orbital configuration 
and  with energy in terms of Racah parameters
\cite{Griffith1964} $A+4B \equiv U- 3J^c_H$. This leads to 
$r^c_1 = \frac{1}{1-3J^c_H/U}$.

(ii) The low spin state ${}^1T_1$ on copper sites which involves $d^2 = c^1 a^1$ orbital configuration 
and  with energy in terms of Racah parameters
\cite{Griffith1964} $A+4B +2C \equiv U- J^c_H$. This leads to 
$r^c_2 = \frac{1}{1-J^c_H/U}$.

(iii) The high spin state ${}^3T_1$ on oxygen sites which involves $p^2 = x^1 y^1$ orbital configuration 
and  with energy in terms of Racah parameters
\cite{Griffith1964} $A_o-5B_o \equiv U_p- 3J^p_H$. This leads to 
$R^c_1 = \frac{2U}{\Delta_x+\Delta_y+U_p(1-3J^p_H/U_p)}$.

(iv) The low spin state ${}^1T_2$ on oxygen sites which involves $p^2 = x^1 y^1$ orbital configuration 
and  with energy in terms of Racah parameters
\cite{Griffith1964} $A_o+B_o+2C_o \equiv U_p- J^p_H$. This leads to
$R^c_2 = \frac{2U}{\Delta_x+\Delta_y+U_p(1-J^p_H/U_p)}$.

\subsection{Remarks on the derivation and parameters}
\label{sec:0c}

Firstly, we would like to remark that the physics of superexchange interactions
is very similar in both `orbital exchange' cases discussed above [cf. Eq. (\ref{eq:hb}) and Eq. (\ref{eq:hc})].
Thus, the main (quantitative) difference between these two cases originates in slightly 
renormalized model parameters. Secondly, we should comment on the 
superexchange paths which, due to their small relative contribution
to the low energy Hamiltonian, are neglected in the above derivation:

(i) There is a finite probability that e.g. the $c^1a^1$ (i.e. $t^1_{2g} e^1_g$) configuration in the intermediate state
${}^3T_1$ decays into a $t^2_{2g}$ configuration. However, such process can be neglected, since this means that in the final state of the superexchange process
we would then be left with a transition to a higher energy sector: starting from the initial state with one hole in a $t_{2g}$ configuration
we would end with two holes in the $t_{2g}$ configuration in the final state of the superexchange process and this would cost the energy $\sim \bar{\varepsilon}_b$ or
$\sim \bar{\varepsilon}_c$; the latter energies are typically much larger than the scales of the superexchange interactions.

(ii) We also neglect intermediate states of the kind ${}^1A_1$ or ${}^1E$ or ${}^3A_2$, since
they all require transitions to the higher energy sector  $\sim \bar{\varepsilon}_b$ or
$\sim \bar{\varepsilon}_c$.

(iii) Finally, from the above structure one can see that it is impossible to
have a `mixing' between the $t_{2g}$ orbital excitations, i.e. to have transitions between the 
states with e.g. $a^1_ic^1_{i+1}$ configuration and e.g. $b^1_i a^1_{i+1}$ configuration
-- this is due to: (a) the flavor-conserving hoppings between the $t_{2g}$ orbitals, 
and (b) no on-site hopping [cf. Eq. (\ref{eq:ct})] allowing for a transition from $c^1a^1$ to $b^1a^1$ state, cf. Table~A26 from Ref.~\onlinecite{Griffith1964}.

Finally, let us note that all of the parameters of the spin-orbital model
directly follow from the charge transfer model parameters. Their values are shown in Table~\ref{tab:1}.

\section{Orbiton spectral functions}
\label{sec:1}
Our main purpose is to calculate the orbiton dispersion, which follows from
the two orbiton spectral functions
\begin{align} \label{eq:spectralbl}
A_b(k, \omega)\equiv &\frac{1}{\pi} \lim_{\eta \rightarrow 0} 
\Im \langle  {0} | \sum_{j \sigma} e^{i k j} f^\dag_{j a \sigma} f_{j b \sigma} \nonumber \\
& \times \frac{1}{\omega + E_{{0}}  -\mathcal{{H}} - i \eta }  \sum_{j \sigma} e^{i k j} f^\dag_{j b \sigma} f_{j a \sigma} | {0} \rangle,
\end{align}
and
\begin{align} \label{eq:spectralcl}
A_c(k, \omega) \equiv &\frac{1}{\pi} \lim_{\eta \rightarrow 0} 
\Im \langle  {0} | \sum_{j \sigma} e^{i k j} f^\dag_{j a \sigma} f_{j c \sigma} \nonumber \\
& \times \frac{1}{\omega + E_{{0}}  -\mathcal{{H}} - i \eta }  \sum_{j \sigma} e^{i k j} f^\dag_{j c \sigma} f_{j a \sigma} | {0} \rangle,
\end{align}
where $|0\rangle$ is the ground state of the charge transfer Hamiltonian $\mathcal{H}$, Eq. (\ref{eq:ct}), with energy $E_0$.

In what follows we will concentrate on the low energy version of the charge transfer Hamiltonian,
i.e. the spin-orbital Hamiltonian Eq. (\ref{eq:hspinorb}). Thus, we express
the above formulae for orbiton spectral functions in terms of the orbital psuedospinon operators acting
in the restricted Hilbert space of the spin-orbital Hamiltonian without double occupancies\cite{notegap}:
\begin{align} \label{eq:spectralbl_v2}
A_b(k, \omega)= &\frac{1}{\pi} \lim_{\eta \rightarrow 0} 
\Im \langle  \bar{0} | \sigma_k 
\frac{1}{\omega + E_{\bar{0}}  -\bar{\mathcal{H}} - i \eta }   \sigma_k^\dag | \bar{0} \rangle,
\end{align}
\begin{align} \label{eq:spectralcl_v2}
A_c(k, \omega)= &\frac{1}{\pi} \lim_{\eta \rightarrow 0} 
\Im \langle  \bar{0} | \tau_k
 \frac{1}{\omega + E_{\bar{0}}  -\bar{\mathcal{H}} - i \eta }  \tau_k^\dag | \bar{0} \rangle,
\end{align}
where $|\bar{0}\rangle$ is the ground state of the spin-orbital Hamiltonian $\bar{\mathcal{H}}$ with energy $E_{\bar{0}}$.
It is now easy to verify that, for the realistic regime of parameters defined in Table~\ref{tab:1}, 
the ground state is insulating, ferroorbital (FO) i.e. only orbital $a$ is occupied, 
and antiferromagnetic (AF) (due to its 1D nature and lack of long range order called `quantum' AF in what follows).
This is because the energy cost of populating $b$ or $c$ orbital states $\bar{\varepsilon}_b$ and $\bar{\varepsilon}_c$
is much larger than hopping $t_n$ (cf. Table~\ref{tab:1}). Thus, it is only the
spin Heisenberg Hamiltonian $\bar{\mathcal{H}}_a$ which dictates what is the spin ground state (which is always AF 
for any positive $J_1$ and $R$, cf. Table~\ref{tab:1}).

In the following subsections we calculate the orbiton spectral functions, 
Eqs. (\ref{eq:spectralbl_v2})-(\ref{eq:spectralcl_v2}): firstly by mapping
them onto the spectral functions of the effective $t$--$J$ model problems 
and then by solving these simplified problems numerically. While this
method is not entirely exact, it gives far better approximation of 
the actual spectral function than the commonly used linear
orbital wave approximation, cf. part 1 and part 3 of Appendix~\ref{sec:low}.

\subsection{Mapping onto the effective $t$--$J$ models }
\label{sec:1d}

%
\begin{figure}[t!]
\centering
   \includegraphics[width=1.0\columnwidth]{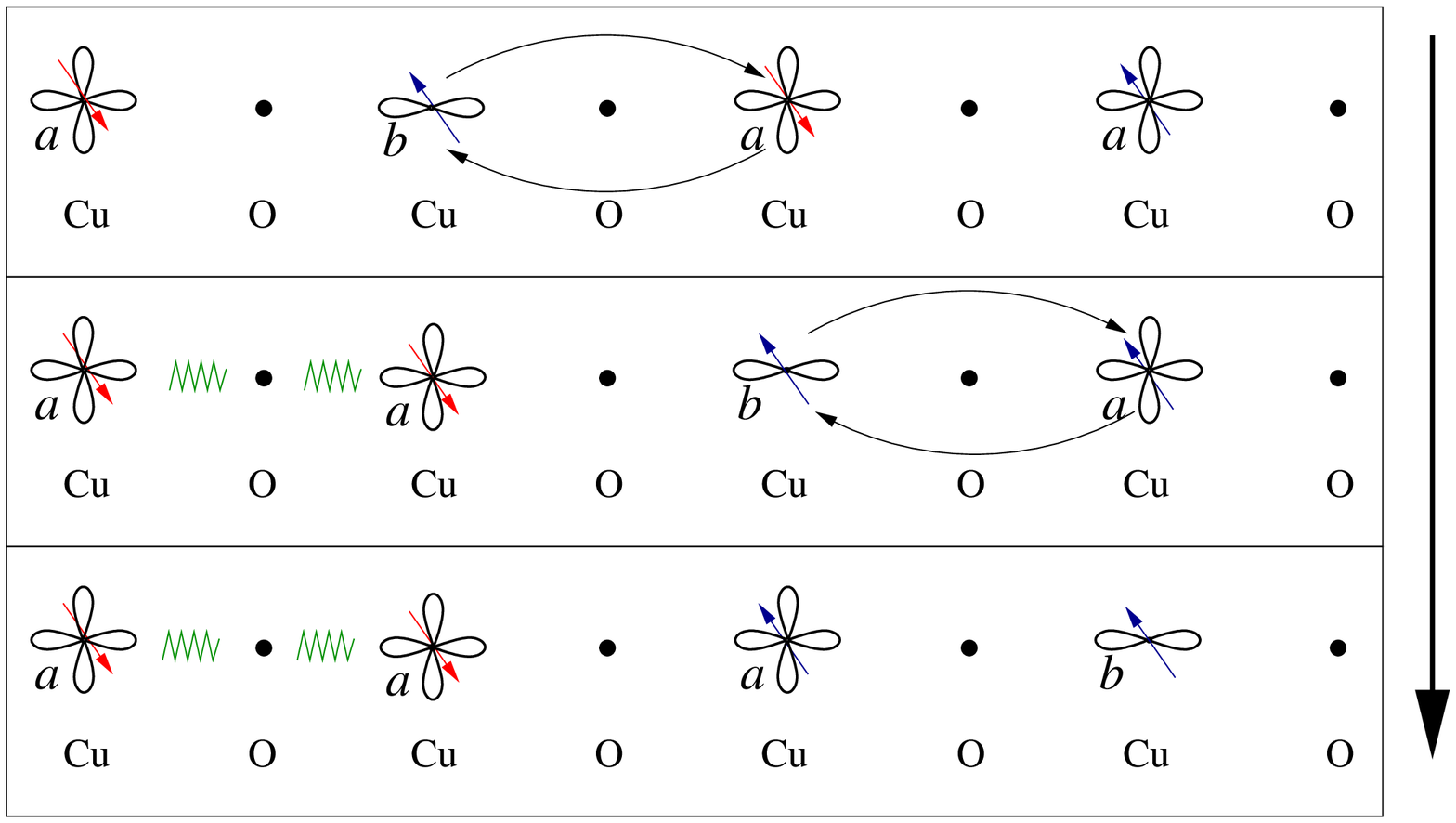}
\caption{(Color online) Schematic view of the propagation of
the orbiton in the spin-orbital separation scenario: the orbiton hops to the neighboring site
(top panel) and initially 
excites a single spinon (middle panel) but then separates from the spinon
and can freely travel in the 1D AF (bottom panel).
}
\label{fig:propSOS}
\end{figure}

{\it Mapping for the $b$ orbiton case.---}
To address the issues mentioned above,
we rewrite Eq. (\ref{eq:spectralbl_v2}) in the following way:
\begin{align} \label{eq:spectralbl_v3}
&A_b(k, \omega)= \frac{1}{\pi} \lim_{\eta \rightarrow 0} 
\Im \Big\langle  \bar{0} \Big| \sum_{j} e^{ikj} \sigma_j \Big(\frac12-S^z_j\Big) \nonumber\\
& \times \frac{1}{\omega + E_{\bar{0}}  -\bar{\mathcal{H}}- i \eta }   \sum_{j} e^{ikj} \sigma^\dag_j \Big(\frac12-S^z_j\Big)  \Big| \bar{0} \Big\rangle \nonumber \\
 &+\frac{1}{\pi} \lim_{\eta \rightarrow 0} 
\Im \Big\langle  \bar{0} \Big| \sum_{j} e^{ikj} \sigma_j \Big(\frac12+S^z_j\Big) \nonumber\\
& \times \frac{1}{\omega + E_{\bar{0}}  -\bar{\mathcal{H}}- i \eta }   \sum_{j} e^{ikj} \sigma^\dag_j \Big(\frac12+S^z_j\Big)  \Big| \bar{0} \Big\rangle.
\end{align}
Here we used an approximation that 
the spectral function for an orbiton is only nonzero when the spin of the hole in the excited orbital
is conserved [i.e. we assumed that the spectral functions for orbiton, which contains the `cross terms' of the 
kind $ \propto (\frac12+S^z_j) (\frac12-S^z_j)$, can be neglected]. In fact, this approximation
amounts to neglecting the process which describes orbiton propagation 
with an additional spin flip, see part 3 of Appendix \ref{sec:low} for justification that such process
has relatively small amplitude and can be neglected. Furthermore, due to the SU(2) spin
invariance of both the ground state $|\bar{0}\rangle$ and of the Hamiltonian $\bar{\mathcal{H}}$,
the two contributions to the spectral function [as written on the right hand side of Eq. (\ref{eq:spectralbl_v3})] are equal, i.e.
we can write
\begin{align} \label{eq:spectralbl_v4}
&A_b(k, \omega)= \frac{2}{\pi} \lim_{\eta \rightarrow 0} 
\Im \Big\langle  \bar{0} \Big| \sum_{j} e^{ikj} \sigma_j \Big(\frac12-S^z_j\Big) \nonumber\\
& \times \frac{1}{\omega + E_{\bar{0}}  -\bar{\mathcal{H}}- i \eta }   \sum_{j} e^{ikj} \sigma^\dag_j \Big(\frac12-S^z_j\Big)  \Big| \bar{0} \Big\rangle. 
\end{align}
 
Next we introduce fermions (to be called spinons) $\alpha$ through the Jordan-Wigner (JW) transformation for spins
and fermions $\beta$ (to be called pseudospinons) through the JW transformation for pseudospins, cf. Ref.~\onlinecite{Wohlfeld2011}.
We define
\begin{align} \label{eq:slambdaneg}
 S^+_j  &= \exp \Big( - i \pi \sum_{n=1,..., j-1} {Q}_n \Big)  \alpha^\dag_j, \nonumber \\
 S^-_j  &=  \alpha_j \exp \Big(  i \pi \sum_{n=1,..., j-1} {Q}_n \Big), \nonumber \\
 S^z_j &= n_{j \alpha} - \frac12.
\end{align}
where ${Q}_n = \alpha^\dag_n \alpha_n$ and $\alpha_n$ are fermions. 
Besides, we define the orbital fermionic operators $\beta$ as:
\begin {align}
\sigma^+_j &= \exp \Big( - i \pi \sum_{n=1,..., j-1} \bar{Q}_n \Big) \beta^\dag_j, \nonumber \\
\sigma^-_j & = \beta_j \exp \Big(  i \pi \sum_{n=1,..., j-1} \bar{Q}_n \Big), \nonumber \\
\sigma^z_j & = n_{j \beta} - \frac12,
\end{align}
where $\bar{Q}_n = \beta^\dag_n \beta_n$ and $\beta_n$ are fermions.

It turns out that when calculating the orbiton spectral
function Eq.~(\ref{eq:spectralbl_v4}) with spins and pseudospins expressed in terms of the JW fermions
following the above transformation, a pseudospinon and spinon are not present
on the same site, cf. Ref.~\onlinecite{Wohlfeld2011}. In other words, 
we have a constraint
\begin{align}\label{eq:constraint}
\forall_i \Large( \beta^\dag_i \beta_i + \alpha^\dag_i \alpha_i \Large) \leq 1,
\end{align}
since otherwise the right hand side of the 
spectral function in Eq. (\ref{eq:spectralbl_v4}) is zero because $\sigma^\dag_j(\frac12-S^z_j) = \beta^\dag_j(1-n_{j \alpha})$.
The physical understanding of this phenomenon is as follows: suppose
one promotes a hole with spin down to the $b$ orbital at site $i$, 
which means that we have no pseudospinon and spinon at this site.
Now, this pseudospinon can move only via such processes
which do not flip the spin of the hole in the $b$ orbital, i.e. we prohibit creating spinon and pseudospinon at the same
site [cf. part 3 of Appendix~\ref{sec:low}].

Altogether this means that while rewriting the low energy Hamiltonian Eqs. (\ref{eq:new1}), (\ref{eq:h1}) and (\ref{eq:hb}) in terms of fermions
$\alpha$ and $\beta$ we can skip all the terms which contain the pseudospinon and 
spinon at the same site.  We arrive at the following Hamiltonian
\begin{align}
H^{ab}_{\rm JW}\equiv  H^0_{\rm JW}+ H^a_{\rm JW} + H^b_{\rm JW}
\end{align}
with
\begin{align}
H^0_{\rm JW} =\bar{\varepsilon}_b \sum_i \beta^\dag_i \beta_i,
\end{align}
\begin{align}
 H^a_{\rm JW}=& J_1 (1+R) \sum_i (1-n_{i \beta}) \Big[ \frac12 (\alpha^\dag_i \alpha_{i+1} + h.c.)  \nonumber \\
&-\frac12 n_{i \alpha} -\frac12 n_{i+1 \alpha} + n_{i \alpha} n_{i+1 \alpha} \Big]
(1-n_{i+1 \beta}),
\end{align}
(with the pseudospinon operators originating in the projection operators $\mathcal{P}_{i, i+1}$) and 
\begin{align}
 H^b_{\rm JW}\! =&\! -\frac14 (R^b_1+r^b_1+R^b_2+r^b_2) J^b_{12} \sum_i \big(\alpha^\dag_i\beta_i\beta^\dag_{i+1} \alpha_{i+1}  + h.c. \big) \nonumber \\
& - \frac12 (R^b_1+r^b_1) J^b_{12} \sum_i \big(\beta_i\beta^\dag_{i+1} + h.c.\big) \nonumber \\
 &- \Big( R^b_1 J^b_{12} + r^b_1 \frac{J_1+J^b_2}{2} \Big) \sum_i \beta^\dag_i \beta_i,
\end{align}
where in addition we assumed that only one pseudospinon in the bulk is present (which corresponds to the FO ground state
with one orbital excitation)
and we skipped the terms $n_{i \alpha} n_{i+1 \beta}+n_{i+1 \alpha} n_{i \beta}$,
as for realistic value of $J^b_H$ and $J_H^p$ (cf. Table \ref{tab:1}) they are of the order of 10\%-20\% 
of the value of the hopping $t_b$ [see Eq. (\ref{eq:tb}) below and Table \ref{tab:1}].

Next, we perform a transformation that connects the above derived Hamiltonian with the effective $t$--$J$ model.
Thus, we introduce the auxiliary fermions $\tilde{p}_{i \sigma}$ acting
in a Hilbert space without double occupancies [we have checked that the operators $\tilde{p}_{i \sigma}$ fulfill the appropriate commutation rules (cf. Ref. \onlinecite{Martinez1991})]:
\begin{align}\label{eq:anothertrafo}
\tilde{p}_{j \uparrow} &= \beta^\dag_j, \nonumber \\
\tilde{p}_{j \downarrow} &= \beta^\dag_j \alpha_j \exp \Big( i \pi \sum_{n=1,..., j-1} {{Q}}_n \Big),
\end{align}
where again ${{Q}}_n = \alpha^\dag_n \alpha_n$. Besides, we introduce back spin operators ${\bf S}$ following Eq. (\ref{eq:slambdaneg}).
Thus, we obtain
\begin{align}
H^{ab}_{t-J} \equiv& H^0_{t-J}+ H^a_{t-J}+H^b_{t-J}= - t_b \sum_{ j, \sigma} (\tilde{p}^\dag_{j \sigma} \tilde{p}_{j+1 \sigma} + h.c.)
\nonumber \\ &+ J \sum_j \Big( {\bf S}_j {\bf S}_{j+1} - \frac14 \tilde{n}_j \tilde{n}_{j+1} \Big)
- E^h_b \sum_j \tilde{n}_j,
\end{align}
where the parameters are defined as 
\begin{align}
 t_b & \equiv  \frac18 (3 R^b_1 +  R^b_2  + 3 r^b_1 + r^b_2) J^b_{12},\label{eq:tb}  \\
 J & \equiv  J_1 (1+R),  \\
 E^h_b & \equiv  \bar{\varepsilon}_b -(R^b_1 J^b_{12} + r^b_1 \frac{J_1+J^b_2}{2}) ,
\end{align}
and we furthermore neglected the difference between $ t_{b \downarrow} \equiv \frac14 ( R^b_1 +  R^b_2  +  r^b_1 + r^b_2) J^b_{12} $
and $ t_{b \uparrow} \equiv \frac12 ( R^b_1 +   r^b_1) J^b_{12} $ hopping element.
Note, however that: (i) this difference is of ca. 10 \% for realistic parameters from Table~\ref{tab:1}
and therefore can be neglected to simplify the calculations,
(ii) keeping this difference while at the same time neglecting the possibility of orbiton propagation with an additional spin flip
(the so-called B1 process in part 3 of Appendix~\ref{sec:low})
violates the SU(2) spin symmetry of the original Hamiltonian,
and finally (iii) we have verified that {\it including} this difference not only does not lead
to qualitatively different RIXS cross section but also the quantitative changes
are negligible.

As the last step we express also the $b$ orbiton spectral function, Eq. (\ref{eq:spectralbl_v4}),
in the $t$--$J$ model language. Using the same transformations as for the Hamiltonian
above we obtain
\begin{align}\label{eq:tbl2}
A_b (k, \omega)= \frac{2}{\pi} \lim_{\eta \rightarrow 0} 
\Im\langle \Phi |  \tilde{p}^\dag_{k \uparrow} \frac{ 1    }{\omega + E_{{\Phi}} - H^{ab}_{t-J} - i \eta } 
 \tilde{p}_{k \uparrow} | \Phi \rangle,
\end{align}
here $|\Phi \rangle$ is the ground state of $H^{ab}_{t-J}$ at half-filling with
energy $E_\Phi$ (i.e. is a 1D quantum AF). Let us note
that the $t$--$J$ model spectral function does not depend on the spin $\sigma$
of the fermion $\tilde{p}_{k \sigma}$ which is consistent with the fact that the choice
of spin $\sigma$ in Eq. (\ref{eq:anothertrafo}) was arbitrary.  

{\it Mapping for the $c$ orbiton case.---} Following the same steps
as for the $b$ orbiton spectral function we obtain the effective
$t$-$J$ Hamiltonian
\begin{align}
H^{ac}_{t-J} \equiv& H^0_{t-J} + H^a_{t-J}+H^c_{t-J}= - t_c \sum_{ j, \sigma} (\tilde{p}^\dag_{j \sigma} \tilde{p}_{j+1 \sigma} + h.c.)
\nonumber \\ &+ J \sum_j  \Big( {\bf S}_j {\bf S}_{j+1} - \frac14 \tilde{n}_j \tilde{n}_{j+1} \Big)
- E^h_c \sum_j \tilde{n}_j,
\end{align}
where the parameters are defined as 
\begin{align}
 t_c & \equiv  \frac18 (3 R^c_1 +  R^c_2  + 3 r^c_1 + r^c_2) J^c_{12},\label{eq:tcc}  \\
 J & \equiv  J_1 (1+R),  \\
 E^h_c & \equiv  \bar{\varepsilon}_c -(R^c_1 J^b_{12} + r^c_1 \frac{J_1+J^c_2}{2}).
\end{align}
The spectral function is then defined as
\begin{align}\label{eq:tcl2}
A_c (k, \omega)= \frac{2}{\pi} \lim_{\eta \rightarrow 0} 
\Im\langle \Phi |  \tilde{p}^\dag_{k \uparrow} \frac{ 1    }{\omega + E_{{\Phi}} - H^{ac}_{t-J} - i \eta } 
 \tilde{p}_{k \uparrow} | \Phi \rangle.
\end{align}

{\it Parameters after the mapping.---} 
As shown above, the parameters $t_b$, $t_c$, and $J$ in the effective $t$--$J$ model
are expressed in terms of the spin-orbital model parameters from the middle column of Table \ref{tab:1}. Thus they can
be easily calculated and their precise values are given in the right column of Table \ref{tab:1}.
On the other hand, while the energies of the on-site orbital excitations $E^h_b$ and  $E^h_c$
also follow from these parameters, in order to stay in line with Ref.~\onlinecite{Schlappa2012}, 
we directly estimate them following the {\it ab-initio} quantum chemistry calculations
on three CuO$_3$ plaquettes in Sr$_2$CuO$_3$, cf. Ref.~\onlinecite{Schlappa2012}.
Note that these {\it ab-initio} calculations are performed for ferromagnetic chain and hence 
they are `well-suited' to our needs, since we have $E^h_b \simeq E_b$ ($E^h_c \simeq E_c$)
where $E_b$ ($E_c$) is defined as the on-site cost of a $b$ ($c$) orbital excitation in a ferromagnetic environment.
Let us note that both methods lead to rather similar results, i.e. estimating
$E^h_b$ and $E^h_c$ directly from the spin-orbital model parameters given in Table~\ref{tab:1}
would lead to similar values as the reported here {\it ab-initio} values.

\subsection{Spin-orbital separation and numerical results}

Altogether, we see that we managed to map the spin-orbital
problem with an FO and AF ground state and one excitation in the $b$  or $c$ orbital 
onto an effective $t$--$J$ model with an AF ground state and one empty site (`hole') 
without a spin (cf. Ref. \onlinecite{Wohlfeld2011} and also Ref. \onlinecite{Bouillot2011}
which also shows a mapping of a spin-like problem onto an effective $t$--$J$ model). 
As the latter problem is well-known~\cite{Kim1996}, 
even before calculating the spectral function, we can draw an interesting
conclusion: The $t$--$J$ model spectral function at half filling, when calculated in 1D,
describes a phenomenon called spin-charge separation. This means that
the `hole' in the 1D AF separates into an independent holon,
which carries charge quantum number, and spinon which carries
spin quantum number. Thus, also the here discussed spin-orbital problem
shows such separation phenomenon -- to be called {\it spin-orbital separation}. 
In fact, this can also be understood by looking at the cartoon
picture in Fig. \ref{fig:propSOS}: (i) the orbiton moves in such a way that the spin
of the hole in this excited orbital is conserved, (ii) this motion introduces
a single defect in the AF ground state (spinon), and (iii) the created spinon
and the `pure' orbiton (~holon in the $t$--$J$ model language) 
can move independently and completely separate~\cite{Wohlfeld2011}. 

\begin{figure}[t!]  
   \includegraphics[width=0.4\textwidth]{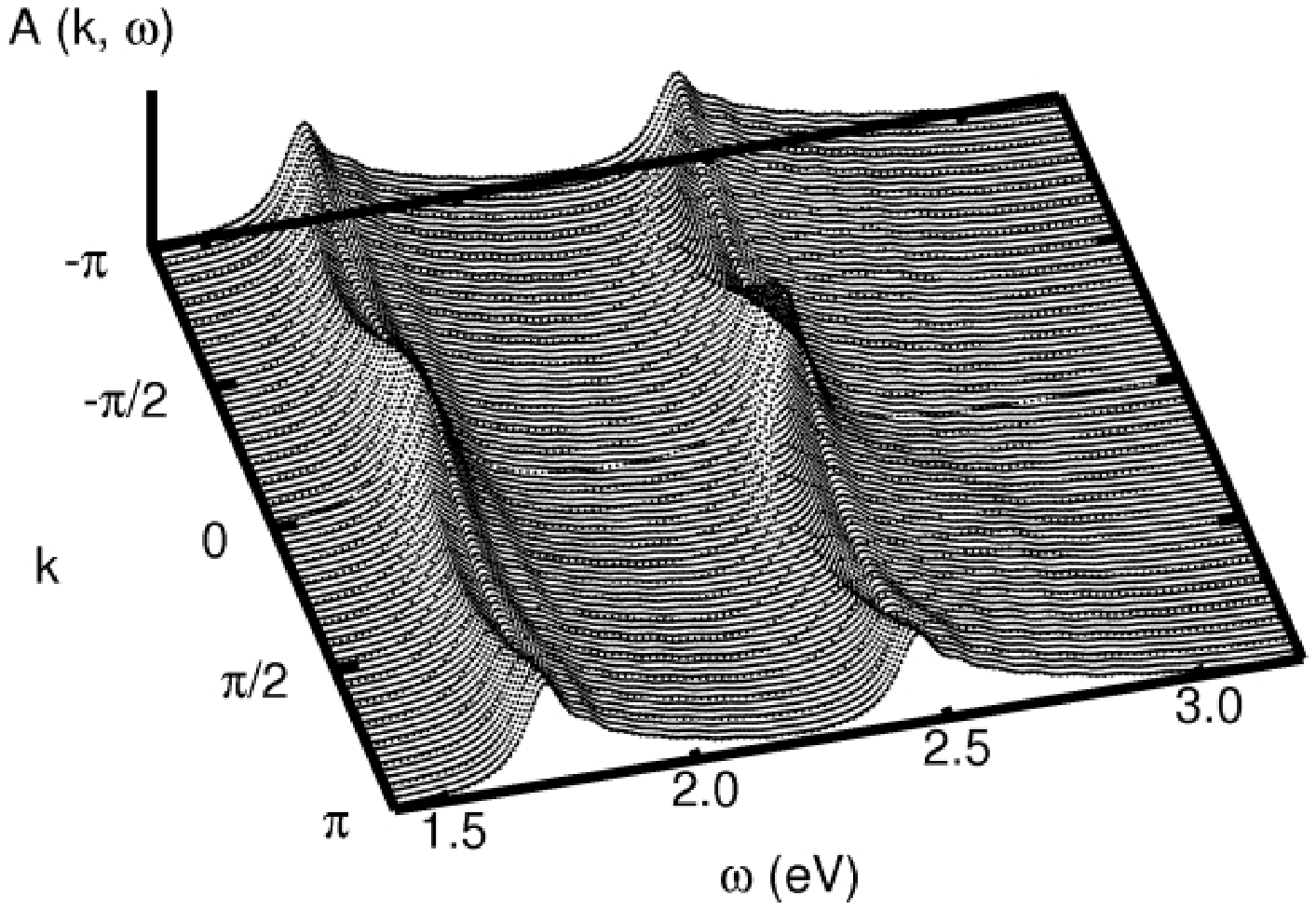}
   \includegraphics[width=0.4\textwidth]{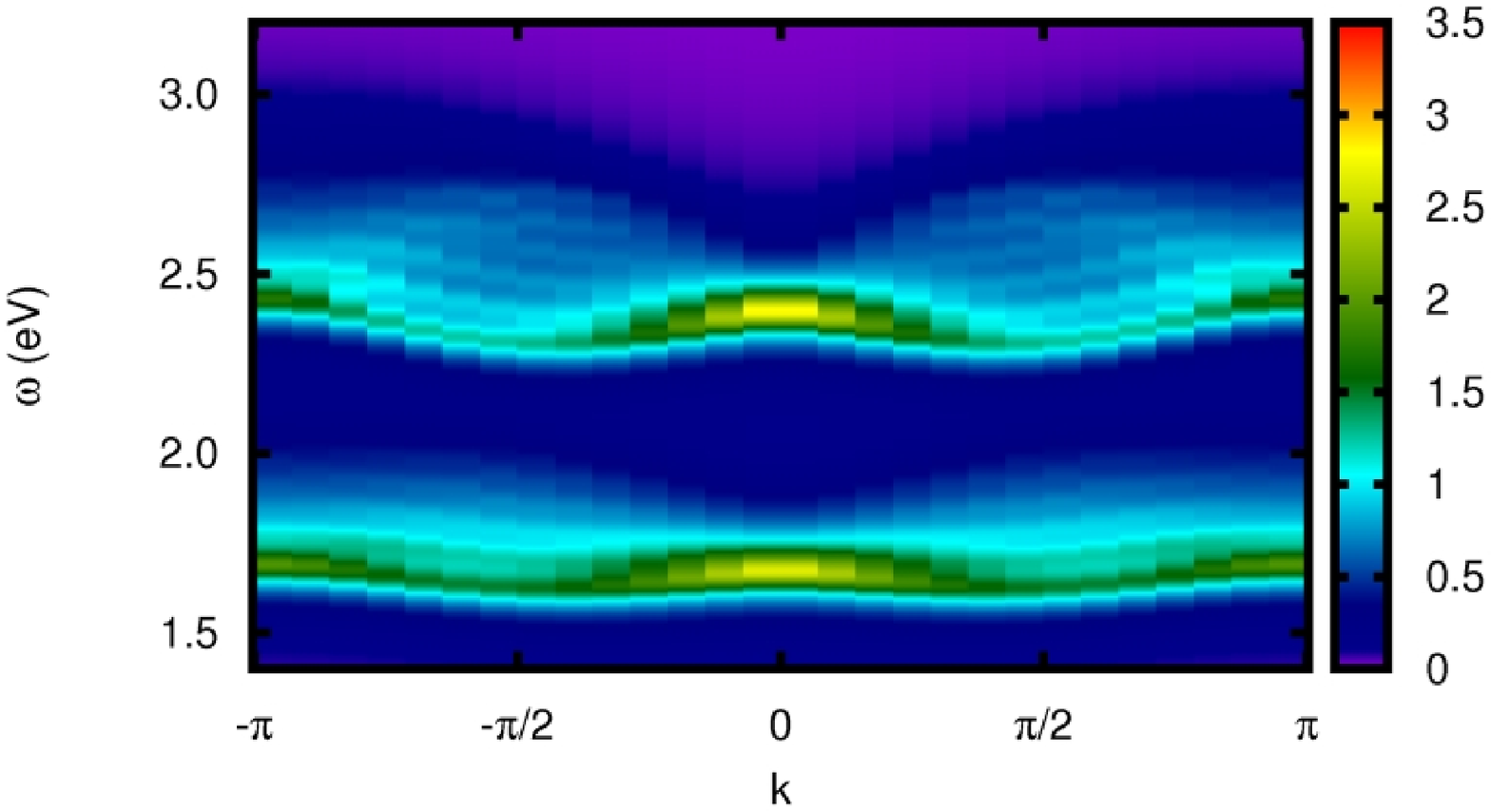}
\caption{(Color online) Spectral function $A_b(k, \omega)+A_c(k, \omega)$ 
as a function of momentum $k$ and energy transfer $\omega$ in the spin-orbital
scenario and calculated using Lanczos exact diagonalization 
on a 28 site chain. Results for broadening 
$\eta = 0.05$ eV which gives
FWHM = 0.1 eV, i.e. the experimental resolution of RIXS
in Sr$_2$CuO$_3$ \cite{Schlappa2012}. .
}
\label{fig:spectrumSOS}
\end{figure}

Nevertheless, i.e. despite the fact that the $t$-$J$ model spectral function is well-known, 
we calculate the spectral functions Eq. (\ref{eq:tbl2}) and Eq. (\ref{eq:tcl2}) using the Lanczos exact 
diagonalization on a 28-site chain separately for each orbiton case. The spectral function for 
the $b$ orbiton [$A_b(k, \omega)$] and $c$ orbiton case  [$A_c(k, \omega)$]  
is shown in Fig. \ref{fig:spectrumSOS}.
The spectrum for each orbiton case consists of a lower lying
orbiton branch with dispersion $\propto t_b$ (or $\propto t_c$), period $\pi$,
and mixed spinon-orbiton excitation bounded from above by 
the edge $\propto \sqrt{J^2+4t^2+4 t J \cos k}$ with $t \equiv t_b$ or 
$t \equiv t_c$ depending on the orbiton under consideration, cf. Ref. \onlinecite{Wohlfeld2011}.
Note that this spectrum is quantitatively (but not qualitatively) different than 
the `usual' spin-charge separation. The latter is `normally' calculated for the case
$J<t$ (whereas in `our' spin-orbital case $J>t$ in the effective $t$-$J$ model)~\cite{Wohlfeld2011}.

\section{RIXS cross section}
\label{sec:2}

\begin{figure}[t!]
\begin{center}
\includegraphics[width=0.25\textwidth]{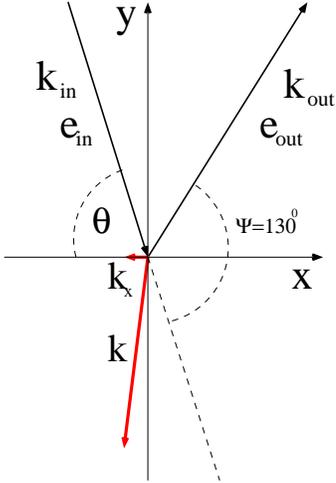}
\end{center} 
\caption{(Color online) Geometry of the RIXS experiment performed on Sr$_2$CuO$_3$ and presented
in Ref. \onlinecite{Schlappa2012} shown here for the $\Psi = 130^0$ scattering angle. 
The momentum transfer $k_x$ is denoted as $k$ in the main text. 
}
\label{fig:geometryRIXS}
\end{figure}
In this section we calculate the RIXS spectra of the orbital excitations in Sr$_2$CuO$_3$.
As it is well-established~\cite{Ishihara2000, Forte2008, Saitoh2001} that RIXS is an excellent 
probe of orbital excitations, the calculations are rather straightforward provided the orbiton
spectral function is known.

Following References~\onlinecite{Haverkort2010, Marra2012}, using the dipole approximation 
and the so-called fast collision approximation~\cite{Luo1993, Groot1998}, 
the RIXS cross section 
for orbital excitations at the Cu$^{2+}$ $L$ edge in the 1D copper oxygen chain reads
\begin{align}\label{eq:i}
I ({ k}, \omega; {\bf e}) = &\frac{1}{\pi} \lim_{\eta \rightarrow 0} 
\Im \langle {0} | T^\dag_S ({ k}, {\bf e})  
\nonumber \\&
\times \frac{1}{\omega + E_{{0}} - \mathcal{{H}} - i \eta } 
T_S ({ k}, {\bf e}) | {0} \rangle,
\end{align}
where  $\omega \equiv \omega_{out} - \omega_{in}$ is the photon energy loss,
${\bf k} \equiv {\bf k}_{in} - {\bf k}_{out}$ is the photon momentum loss 
(with $k \equiv k_x= {\bf k} \cdot {\bf \hat{x}}$ being the momentum loss along the $x$ direction of the copper oxygen chain in the studied case of
Sr$_2$CuO$_3$, cf. Fig. \ref{fig:unitcell}), and $|{0}\rangle$ is the ground state of Hamiltonian $\mathcal{H}$ with energy $E_{{0}}$, see Eq. (\ref{eq:ct}) and cf. Eqs. (\ref{eq:spectralbl})-(\ref{eq:spectralcl}). Finally, $T_S ({k}, {\bf e})$ is the RIXS scattering operator, which depends on the incoming and outgoing photon
polarization ${\bf e}= {\bf e}_{in} {\bf e}_{out}^\dag$ in the RIXS experiment. It reads \cite{Groot1998, Haverkort2010, Marra2012} 
\begin{align}\label{eq:tc}
T_S =&  T_b + T_c +T_d + T_e= 
\sum_{j, \sigma, \sigma'} e^{i k j} 
\big[ 
B_{\sigma, \sigma'} ({\bf e})  f^\dag_{{ j} b \sigma}  f_{{ j} a \sigma'}  \nonumber \\ +  
&C_{\sigma, \sigma'} ({\bf e})   f^\dag_{{ j} c \sigma}  f_{{ j} a \sigma'} 
+  
D_{\sigma, \sigma'} ({\bf e})   f^\dag_{{ j} d \sigma}  f_{{ j} a \sigma'}  \nonumber \\
+ & E_{\sigma, \sigma'} ({\bf e})   f^\dag_{{ j} e \sigma}  f_{{ j} a \sigma'} \big] .
\end{align}
Here: (i) orbital $d = 3d_{yz}$, orbital $e = 3d_{3z^2-r^2}$, and the other orbitals
are defined as in Eq. (\ref{eq:ct}), (ii) operator $f^\dag_{j \alpha \sigma}$  and $f_{j \alpha \sigma}$ is defined as in Eq. (\ref{eq:ct})
with $\alpha \in \{a, b, c, d, e \}$, (iii) $B_{\sigma, \sigma'}$, $C_{\sigma, \sigma'}$, $D_{\sigma, \sigma'}$, and 
$E_{\sigma, \sigma'}$ are complex numbers which define the so-called RIXS matrix elements. 
The latter ones can be easily calculated in the fast collision approximation, see immediately below.

\subsection{RIXS matrix elements}
\label{sec:matrix}
To calculate the above defined RIXS matrix elements in the fast collision approximation, and to be able to compare
the obtained results with the experimental ones reported in Ref. \onlinecite{Schlappa2012},
we assume that: (i) the incoming energy of the photon is tuned to the copper $L_3$ edge, i.e. $\omega_{in} \simeq 930$ eV, and 
thus the wavevector of the incoming photon is $k_{in} \simeq 0.47 1 / \text{\AA}$, (ii) the wavevector at the edge of the Brillouin
zone along the $x$ direction in Sr$_2$CuO$_3$ is $0.805 1/ \text{\AA}$ as the lattice
constant is~\cite{Kojima1997} $3.91 \text{\AA}$,  
(iii) in the ionic picture the ground state
configuration at the copper site is $3d^9$, 
(iv) the  relatively small spin-orbit coupling in
the $3d$ shell can be neglected,
(v) the incoming polarization vector
 ${\bf e}_{in}$ is parallel to the scattering plane and the outgoing polarization vector is not measured, cf. Fig.~\ref{fig:geometryRIXS},
(vi) the scattering plane is the $xy$ plane,i.e. the one in which the copper oxygen
chain lies (which runs along the $x$ direction, see Fig.~\ref{fig:unitcell}), cf. Fig.~\ref{fig:geometryRIXS},
and (vii) the angle between the outgoing and the incoming photon momentum is either $\psi = 90^0$
or $\psi =130^0$, cf. Fig. \ref{fig:geometryRIXS}. The latter defines the two scattering geometries used
in the RIXS experiment reported in Ref. \onlinecite{Schlappa2012}.

Next, we calculate the RIXS matrix elements in three steps: 

Firstly, following {\it inter alia} Ref. \onlinecite{Marra2012},
we express the matrix elements in terms of the different components of the incoming and outgoing polarization vectors
and the spin operator. These expressions can be easily obtained from Fig. 1 of Ref. \onlinecite{Marra2012} and hence
we do not write them here. 

Secondly, we express the incoming and outgoing polarization vectors in terms of the 
angle $\theta$ measured between the momentum of the incoming photon and the $x$ chain direction, cf. Fig. \ref{fig:geometryRIXS}
(note the convention that if $k< 0 $, then $\theta \rightarrow 0$): (i) the vector of the incoming polarization of the photon in 
terms of the angle $\theta$
is ${\bf e}_{in} = [\sin \theta , \cos \theta, 0]$,
(ii) the vector of the outgoing polarization of the photon in terms of the angle $\theta$
is ${\bf e}_{out} = [-\cos \theta , \sin \theta, 0]$ for $\pi$ polarization and
${\bf e}_{out} = [0,0, 1]$ for $\sigma$ polarization in the $\psi= 90^0$ geometry,
and (iii) the vector of the outgoing polarization of the photon in terms of the angle $\theta$
is $ {\bf e}_{out}= [-\cos (\theta-40^0) , \sin (\theta-40^0), 0]$ for $\pi$ polarization and
${\bf e}_{out} = [0,0, 1]$ for $\sigma$ polarization in the $\psi = 130^0$ geometry.

Thirdly, we express the angle $\theta$ in terms of the transferred momentum $k$ along the $x$ direction:
(i) for $\psi = 90^0$ geometry the transferred momentum as a function of angle $\theta \in (0^0, 90^0)$  is
$k \simeq 0.58 \sqrt{2} \pi \sin (\theta-45^0) $ where the distance between the copper sites along the chain
is assumed to be equal to unity, and
(ii) for  $\psi= 130^0$ the angle changes as $\theta \in (0^0, 130^0)$  and 
$k \simeq 1.07 \pi \sin (\theta-65^0) $. 

Altogether, this shows how to calculate the RIXS matrix elements
for orbital excitations and that the latter effectively becomes a function of transferred momentum $k$ and
energy $\omega$. Therefore, in what follows, we simplify notation and write $ I ({ k}, \omega; {\bf e}) \rightarrow  I ({ k}, \omega) $.

\subsection{Numerical results} 
\begin{figure*}[t!]
   \includegraphics[width=0.4\textwidth]{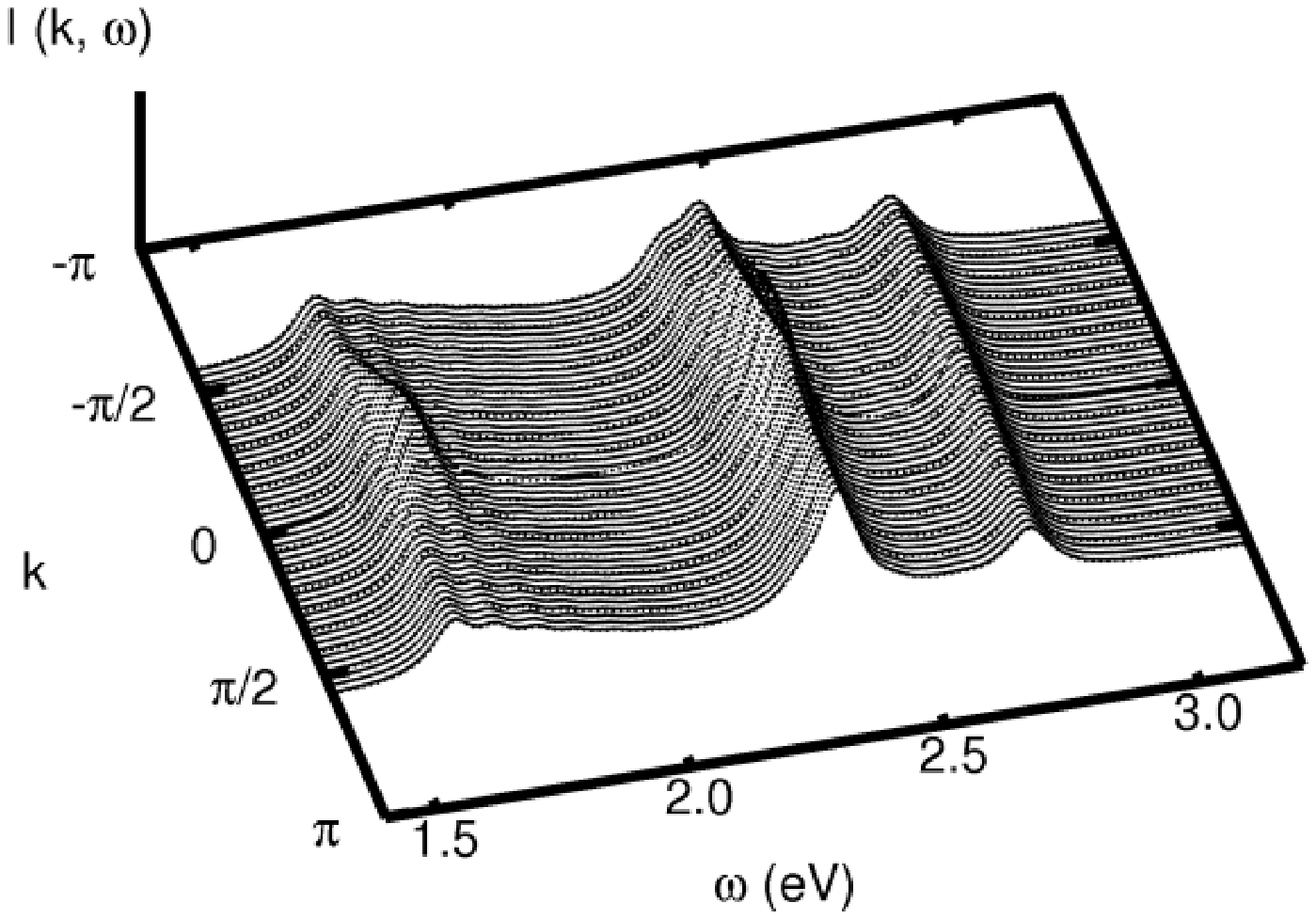}
   \includegraphics[width=0.4\textwidth]{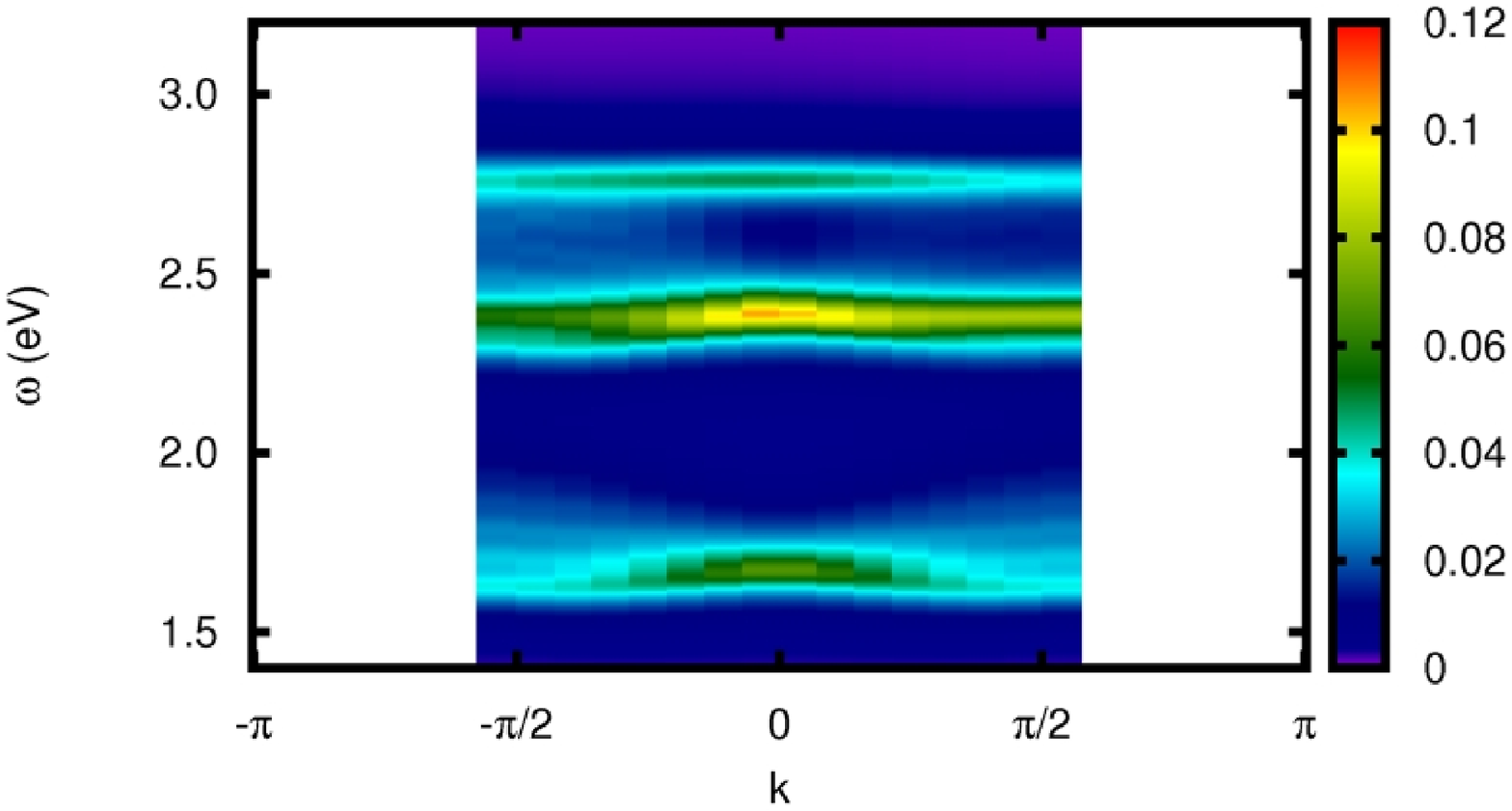}
   \includegraphics[width=0.4\textwidth]{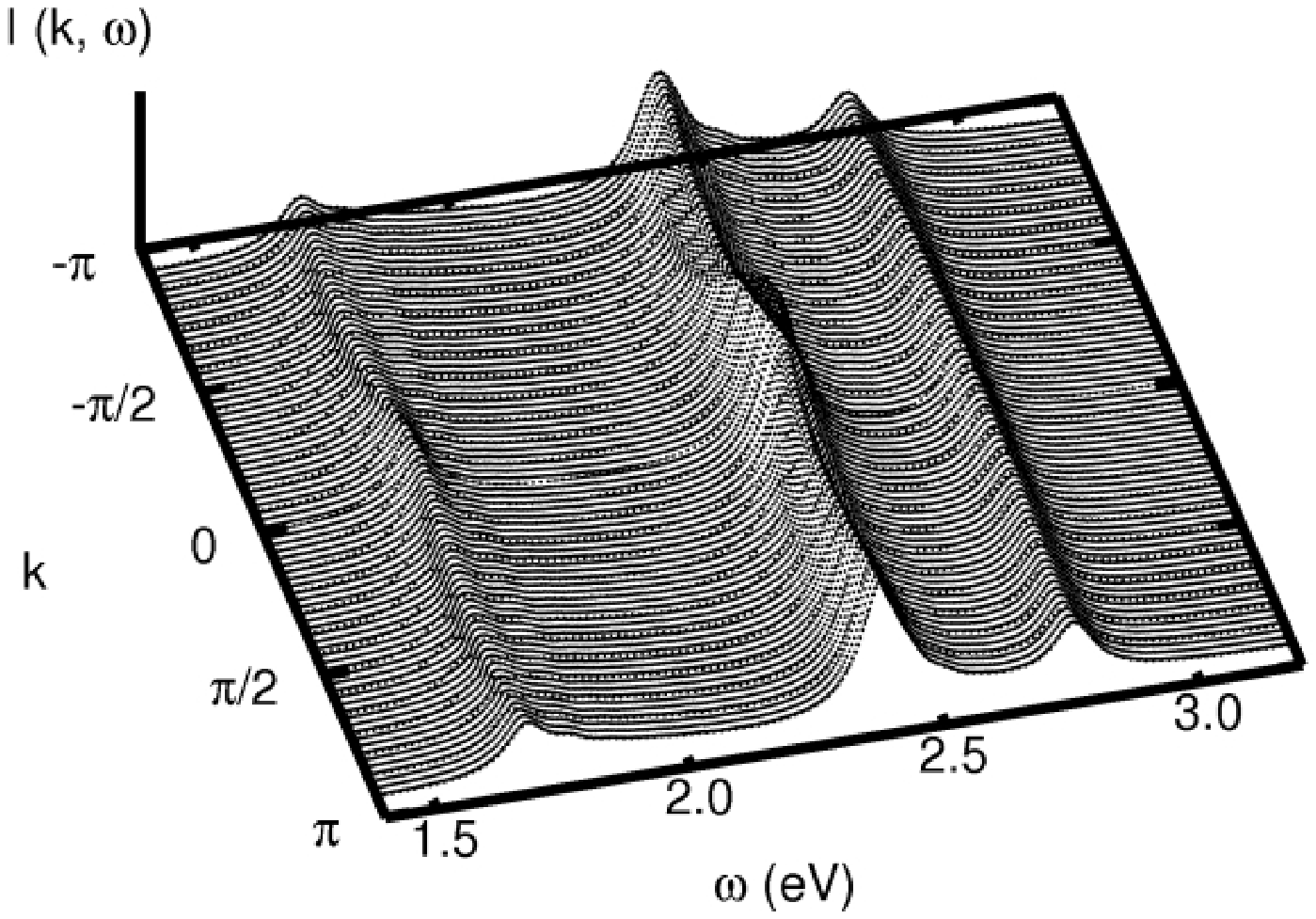}
   \includegraphics[width=0.4\textwidth]{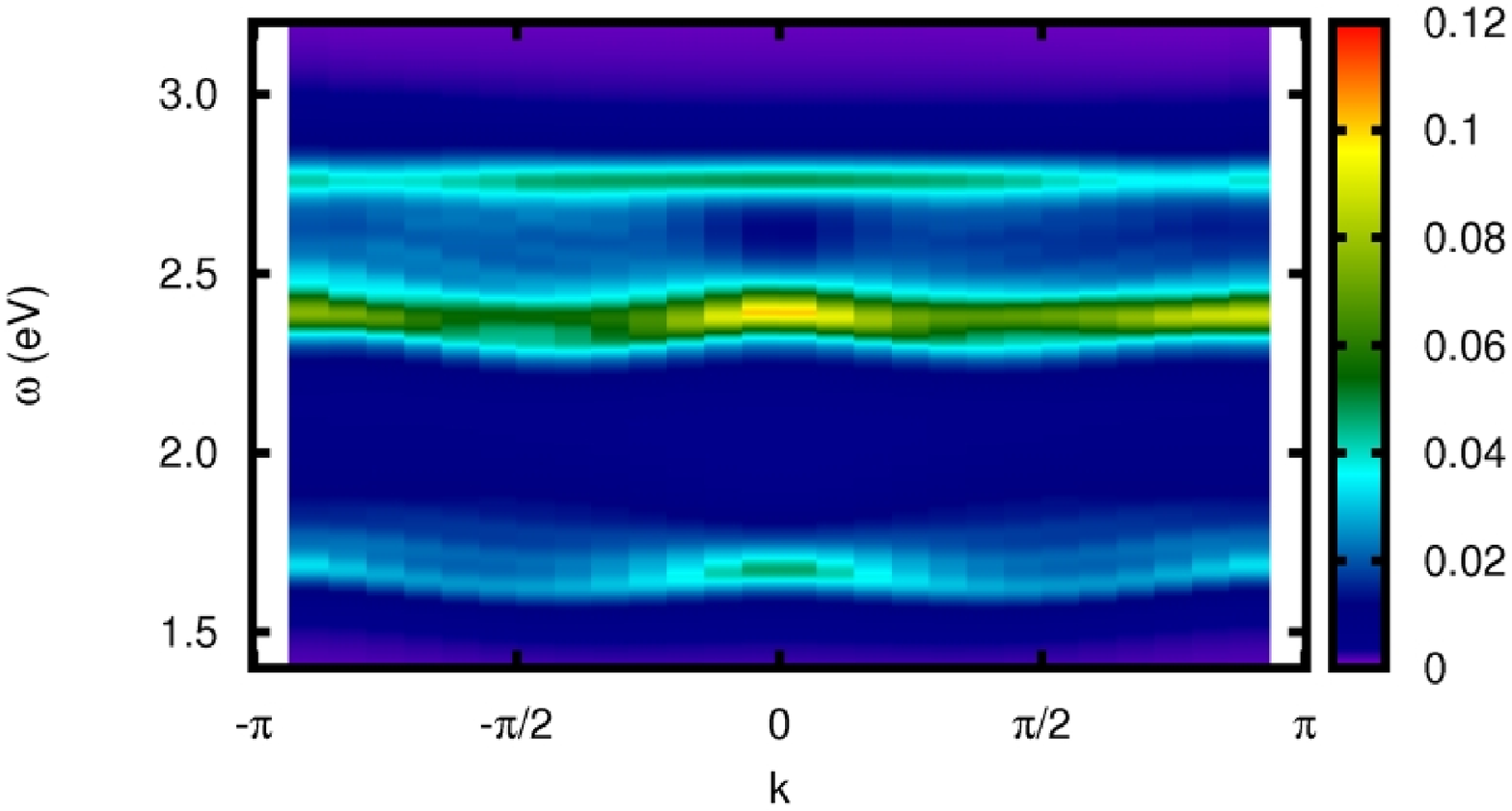}
\caption{(Color online) RIXS cross section for $\psi = 90^0$ ($\psi =130^0$) scattering 
geometry as calculated in the spin-orbital separation scenario and convoluted
with the results from the local model, Fig.~\ref{fig:localRIXS}, 
on the top (bottom) panel.
Left (right) panels show line (color map) spectra.
Results for broadening $\eta =0.05$ eV (cf.
caption of Fig. \ref{fig:spectrumSOS}).
}
\label{fig:SOSRIXS}
\end{figure*}

We first express the RIXS cross section for the $b$ and $c$ orbital excitations
in terms of the previously calculated (see Sec.~\ref{sec:1}) $b$ and $c$ orbiton
spectral functions. In order to do so, we use 
the fact that according the the analysis in Sec.~\ref{sec:1d} the spin of the hole in the excited orbital
does not change during the orbiton propagation process.
Thus, for e.g. only the $b$ part of the RIXS cross section we can write:
\begin{align}
I_b (k,\omega) & = \frac{1}{\pi} \lim_{\eta \rightarrow 0} 
\Im \langle {0} | 
\sum_{\sigma_1, j}B^*_{\sigma_1, \uparrow}  e^{i k j} f^\dag_{j a \sigma_1} f_{j c \uparrow}  
\nonumber \\
&\times \frac{ 1   }{\omega + E_0 -  \mathcal{H}  - i \eta } 
 \sum_{\sigma_2, j}  B_{\uparrow, \sigma_2} e^{i k j} f^\dag_{j c \uparrow} f_{j a \sigma_2}  
 | {0} \rangle 
 \nonumber \\
&+\frac{1}{\pi} \lim_{\eta \rightarrow 0} 
\Im \langle {0} | 
\sum_{\sigma_1, j}B^*_{\sigma_1, \downarrow}  e^{i k j} f^\dag_{j a \sigma_1} f_{j c \downarrow}  
\nonumber \\
&\times \frac{ 1   }{\omega + E_0 -  \mathcal{H}  - i \eta } 
 \sum_{\sigma_2, j}  B_{\downarrow, \sigma_2}  e^{i k j} f^\dag_{j c \downarrow} f_{j a \sigma_2}  
 | {0} \rangle. 
\end{align}
Next, we can employ the transformations used in Sec. \ref{sec:0} and \ref{sec:1d} to map the above
problem first onto a spin-orbital model and then onto an effective $t$-$J$ model problem. We obtain then: 
\begin{align}
I_b({k},\omega) & =\frac{1}{\pi} \lim_{\eta \rightarrow 0} 
\Im \langle \Phi | ( B^*_{\uparrow, \uparrow}\tilde{p}^\dag_{k \uparrow} +  B^*_{\uparrow, \downarrow} \tilde{p}^\dag_{k \downarrow}) \nonumber \\
&\frac{ 1     }{\omega + E_{{\Phi}} - H^{ab}_{t-J} - i \eta} 
( B_{\uparrow, \uparrow} \tilde{p}_{k \uparrow} +  B_{\uparrow, \downarrow} \tilde{p}_{k \downarrow}) | \Phi \rangle \nonumber \\
&+\frac{1}{\pi} \lim_{\eta \rightarrow 0} 
\Im \langle \Phi | ( B^*_{\downarrow, \downarrow} \tilde{p}^\dag_{k \uparrow} +  B^*_{\downarrow, \uparrow} \tilde{p}^\dag_{k \downarrow}) \nonumber \\ 
&\frac{ 1     }{\omega + E_{{\Phi}} - H^{ab}_{t-J} - i \eta} 
( B_{\downarrow, \downarrow} \tilde{p}_{k \uparrow} +  B_{\downarrow, \uparrow} \tilde{p}_{k \downarrow}) | \Phi \rangle.
\end{align}
Finally, the `interference' terms in the above equation cancel due to the identity relations between the RIXS matrix elements
\begin{align}
 B_{\uparrow, \uparrow} B^*_{\downarrow, \uparrow} =  -B^*_{\downarrow, \downarrow} B_{\uparrow, \downarrow}
\end{align}
which leads to:
\begin{align}
&I_b ({ k},\omega)  = 
(|B_{\uparrow, \uparrow}|^2 +  |B_{\downarrow, \uparrow}|^2 )A_b (k, \omega),
\end{align}
where we used that (see Sec. \ref{sec:1d})
\begin{align}
A_b (k, \omega) = & \frac{1}{\pi} \lim_{\eta \rightarrow 0}\Im
\langle \Phi |  \tilde{p}^\dag_{k \uparrow} \frac{ 1     }{\omega + E_{{\Phi}} - H^{ab}_{t-J} - i \eta } 
 \tilde{p}_{k \uparrow} | \Phi \rangle \nonumber \\
 = &
 \frac{1}{\pi} \lim_{\eta \rightarrow 0} \Im
\langle \Phi |  \tilde{p}^\dag_{k \downarrow} \frac{ 1     }{\omega + E_{{\Phi}} - H^{ab}_{t-J} - i \eta } 
 \tilde{p}_{k \downarrow} | \Phi \rangle
 \end{align}
and $A_b (k, \omega)$ was calculated in Sec. \ref{sec:1}.
Employing the same transformation for the $c$ orbiton, we obtain
$I_c ({ k},\omega)  = 
(|C_{\uparrow, \uparrow}|^2 +  |C_{\downarrow, \uparrow}|^2 )A_b (k, \omega)$ 
with $A_c (k, \omega)$ also calculated in Sec. \ref{sec:1}.

Secondly, to complete the RIXS calculations we also have to add the spectra for the dispersionless
excitations to the $d$ and $e$ orbitals. As this task is straightforward, cf. Appendix \ref{sec:2a}, 
we obtain for the total RIXS cross section
\begin{align}\label{eq:final}
I ({ k},\omega)  = &(|B_{\uparrow, \uparrow}|^2 +  |B_{\downarrow, \uparrow}|^2 )A_b (k, \omega) \nonumber\\
&+(|C_{\uparrow, \uparrow}|^2 +  |C_{\downarrow, \uparrow}|^2 )A_c (k, \omega) \nonumber \\
&+(|D_{\uparrow, \uparrow}|^2+|D_{\uparrow, \downarrow}|^2) \delta(\omega - E_d- E_{AF}) \nonumber \\ 
&+(|E_{\uparrow, \uparrow}|^2+|E_{\uparrow, \downarrow}|^2) \delta(\omega - E_e- E_{AF}).
\end{align}
Note that the values of the on-site orbital energies of the dispersionless orbital excitations, $E_d$ and $E_e$, are obtained from the quantum chemistry {\it ab-initio}
calculations for a ferromagnetic chain consisting of three CuO$_3$ plaquettes, cf. Table~\ref{tab:1}.
Since these values are given for a ferromagnetic chain, we have to add the energy cost of a single
spin flip ($E_{AF}$), which is also calculated using the same {\it ab-initio} method, cf. Table~\ref{tab:1} for its precise value. 

{\it Comparison with the experiment.---} 
The RIXS cross section calculated according to Eq.~(\ref{eq:final}) is shown
in Fig.\ref{fig:SOSRIXS}. Comparing this theoretical spectrum against the
experimental one shown in Fig. 4(a) in Ref. \onlinecite{Schlappa2012}
[for the case of the scattering angle $\Psi = 130^0$; a similar agreement is obtained
for the unpublished RIXS experimental results~\cite{privcomm} for the scattering angle $\Psi = 90^0$]
we note the following similarities between the two:
\begin{itemize}
 \item The $c$ orbiton spectrum: both the dispersion and the intensities 
 agree qualitatively and quantitatively; in particular 
 the theoretical spectrum has the largest intensity at $k=0$ 
 momentum which is solely a result of the dispersion originating 
 in the spin-orbital separation scenario (the RIXS local
 matrix elements for the $c$ orbiton are momentum-independent
 in the RIXS geometry of Fig.~\ref{fig:geometryRIXS}).
\item The $b$ orbiton dispersion: the dispersion has the same particular cosine-like
shape with a period $\pi$ and minima at $\pm \pi/2$;
the spectral weights agree qualitatively and quantitatively.
\item `Shadow' (`oval'-like) bands above the $b$ orbiton: the width and shape of the shadow
band is very similar both in the experiment and in theory; spectral weights agree 
relatively well (e.g. larger spectral weights for the negative than for the positive 
momentum transfer).
\end{itemize}

The characteristic spin-orbital
separation spectrum is much better visible for the $b$ orbiton than for the $c$ orbiton.
The reason for this is twofold. Firstly, the overall sensitivity of RIXS to the $b$ orbiton 
excitations is much larger than to the $c$
orbiton due to the chosen geometry of the RIXS experiment. Thus all features
related to the $b$ orbiton are better visible than those related to the $c$ orbiton.
Secondly, even despite this, the spin-orbital separation can be better observed
for the $b$ orbiton case than for the $c$ orbiton, cf. Fig. \ref{fig:spectrumSOS}. This is because
the effective hopping element $t_b$ for the $b$ orbiton is larger than the hopping $t_c$ for the $c$ orbiton, cf.~Table \ref{tab:1},
which is due to: (i) the renormalization $\propto \lambda_c$ of the copper oxygen hopping $t_\pi$ for the hopping from $c$ orbiton 
due to the formation of the bonding and antibonding states with the neighboring oxygens (see App.~\ref{sec:applambda}), 
and (ii) the larger effective charge transfer gap for the $c$ orbital than for the $b$ orbital
(see Table \ref{tab:1}; note that this effective charge transfer gap is defined as the difference in energy between
a particular 3$d$ orbital and the {\it hybridizing} 2$p$ orbital and thus is larger for lower lying 3$d$ orbitals).

The main discrepancy between the experiment and theory is related to the somewhat smaller dispersion in the theoretical calculations
than in the experiment. While there might be several reasons explaining this fact,
let us point two plausible ones. Firstly, the neglected spin-orbit coupling
in the 3$d$ orbitals would mix the $b$ and $d$ (i.e. $xz$ and $yz$) orbital excitations
and would lead to a finite dispersion in the $d$ orbital channel. This would
mean that the present dispersionless $d$ orbital excitation would no longer
`cover' parts of the dispersive $b$ orbiton. Thus, effectively this would lead to large dispersive
feature around the $b$ and $d$ excitation energy. Secondly, the relatively high covalency of the Sr$_2$CuO$_3$ compound,
which is not taken into account in the present derivation, might lead to
the more itinerant character of the system and larger dispersion relation
for the orbiton. 

{\it Comparison with other theoretical calculations.---}
There are actually two other simple approximations which might naively be employed to calculate
the RIXS spectra and which could be compared against the experiment:
(i) the `local model' approximation which assumes that {\it all} orbital
excitations are local, i.e. also both the $b$ and the $c$  orbiton spectral function
do not have any momentum dependence (cf. Appendix \ref{sec:2a}),
and (ii) the one which assumes that the spectral functions of the $b$ and $c$ orbitons are calculated
using the linear orbital wave approximation (cf. part 2 of Appendix \ref{sec:low}).
However, as shown in detail in the above mentioned appendices,
the RIXS spectra calculated using these approximations do not fit the experimental ones.

\section{Discussion and conclusions}
\label{sec:3}

We first list a few alternative scenarios that might lead to the dispersive orbital excitations
in Sr$_2$CuO$_3$ and argue why they do not lead to a plausible explanation of the experimental
results reported in Ref. \onlinecite{Schlappa2012}. At the end of the section we present our conclusions.

\subsection{Alternative scenarios leading to dispersive features in the RIXS spectrum of Sr$_2$CuO$_3$}
{\it Spin-charge separation observed directly.---}
The spin-charge separation where a hole created in a 1D AF decays into
a holon and a spinon can be observed with ARPES: e.g. in SrCuO$_2$~\cite{Kim1996}
or in Sr$_2$CuO$_3$~\cite{Fujisawa1999}. However, not only that holon dispersion is much
larger than the dispersion under consideration in this paper 
(it is of the order of 1.1 eV~\cite{Fujisawa1999}), but also -- what is more important -- 
one cannot directly probe the spin-charge separation in RIXS, since 
in RIXS the total charge is conserved \cite{Kim2004}.

{\it Holon, antiholon and two spinons, i.e. spin-charge separation indirectly---}
Nevertheless, it occurs that there is a possibility to observe spin-charge separation with 
RIXS or EELS in an indirect way. If one transfers a hole from the copper site $i$ either to the neighboring copper
site $i+1$ to form a doubly occupied site {\it or} to the neighboring oxygen plaquette surrounding
the central copper site $i + 1$ to form a Zhang-Rice singlet, then one ends up with one hole in the spin
background on site $i$ and another hole in the spin background on site $i + 1$. 
Next, both of these objects can become mobile and experience the spin-charge separation: the first one can move by decaying to a holon
and a spinon while the second one can move by decaying to an antiholon and a spinon.
This rather complicated scenario was invoked to explain the $K$ edge RIXS spectra in various quasi-1D cuprates
(cf. Refs.~\onlinecite{Kim2004, Hasan2002}) and the EELS spectrum in Sr$_2$CuO$_3$~\cite{Neudert1998}.

However, the common feature of all these experiments is that 
there is a large dispersion (of the order of 1 eV) which has a periodicity of $2\pi$ and a 
minimum at $k=0$. Thus, clearly it is not the spectrum that is observed in the $L$ edge
RIXS experiment in Ref.~\onlinecite{Schlappa2012}. The reason for this is that 
RIXS at $L$ edge is much more sensitive to the on-site excitations on the
copper site than to the intersite charge transfer excitations on the neighbouring 
oxygen or copper sites~\cite{Geck2012}.

{\it Orbital excitations propagating via the $O(2p)$ orbitals.---}
In that case the propagation would entirely happen via the $O(2p)$ orbitals on a kind of
a zigzag chain along the CuO$_4$ plaquettes. This, however, cannot lead to a momentum
dependence in the observed spectrum.

{\it Similar experiments.---} 
One should also compare the here reported theoretical results to the experimental
ones which were discussed in Ref.~\onlinecite{Seo2006}.
There a somewhat similar dispersive feature, as the one discussed
here, was discovered in the RIXS spectra at the $1s \rightarrow 3d$ edge. 
Although this dispersion was attributed
to an orbital excitation, it remained unclear to the authors of that paper how
to correctly interpret this phenomenon. An obvious suggestion is
that the spectrum observed in Ref.~\onlinecite{Seo2006} might be of 
similar origin as the one discussed here:
the dispersion also has a $\pi$ periodicity and is of the order
of $0.2$ eV. However, there are two problems with this scenario: (i) there
is just one dispersive peak but no other dispersive modes
and there is no shoulder peak, (ii) the dispersion is shifted
by $\pi /2 $ in the momentum space. This first problem 
can perhaps be `solved': RIXS at the $1s \rightarrow 3d$ edge involves quadrupolar transitions, 
the RIXS signal is rather weak, and thus it is
possible that one cannot observe all details of the
spectra. However, the second one remains a challenge for theory.
One suggestion might be that the effective `dispersion' that one sees in the spectrum
in Ref.~\onlinecite{Seo2006} is the top part of the `shadow' bands that 
we reported here for the $b$ orbiton -- this would require that the `true'
(lower) $b$ orbiton band is covered by some other excitations in that experiment.

\subsection{Conclusions}

We have considered in detail the origin of the dispersive features observed
in the RIXS spectra of the quasi-1D CuO$_3$ chain in Sr$_2$CuO$_3$~\cite{Schlappa2012}.
We explained that these dispersive features can indeed be attributed to the dispersive orbital excitations (orbitons)
-- which actually were unambiguously observed for the first time. The unexpectedly strong dispersion of these
excitations are not only a result of the relatively strong superexchange interactions in the system
but also due to the fractionalization of the spin and orbital degrees of freedom which, as
shown in this paper and in Refs.~\onlinecite{Wohlfeld2011, Schlappa2012}, 
is possible in this quasi-1D strongly correlated system. 

Finally, one may wonder whether the spin-orbital separation phenomenon 
can also be observed in other transition metal oxides. While we suggest that this should be possible
in most other quasi-1D system which are again mostly cuprates, it is impossible
to observe this phenomenon in 2D and 3D systems, such as La$_2$CuO$_4$, LaMnO$_3$ or LaVO$_3$, 
with long range magnetic order, cf. Ref. \onlinecite{Wohlfeld2012}. However, many theoretical
studies have discussed the nature of the orbital excitations in these systems which leaves large field to be still
explored experimentally (cf. Refs. \onlinecite{Kugel1973, Feiner1997, Ishihara1997, 
vandenBrink1998, Ishihara2000, Oles2000, Ishihara2004, Khaliullin1997, Khaliullin2000, Kikoin2003, Kim2012}).
Furthermore, the mapping of the spin-orbital model into the effective simpler $t$--$J$ model presented here is actually 
valid also in higher dimensions~\cite{Wohlfeld2011}. However, the lack of experimental results, which can verify various 
theories concerning these orbital excitations, means that it remains a challenge both for theory and for experiment to 
explore the nature of the orbital excitations in higher dimensions.

\section{Acknowledgments}
First and foremost we acknowledge very stimulating discussions and common work on this subject
with our theoretical collaborators Maria Daghofer, Liviu Hozoi, and Giniyat Khaliullin as well as
with our experimental collaborators Justina Schlappa, Thorsten Schmit, and Henrik Ronnow. 
We also thank Stefan-Ludwig Drechsler, Andrzej M. Ole\'s, George A. Sawatzky, Michel van Veenendaal, and Victor Yushhankhai for very valuable comments. K.\,W. acknowledges support from the Alexander von Humboldt Foundation, the Polish National 
Science Center (NCN) under Project No. 2012/04/A/ST3/00331, and from the U.S. Department of Energy, 
Materials Sciences and Engineering Division, under Contract No. DE-AC02-76SF00515. 
This research benefited from the RIXS collaboration supported by the 
Computational Materials Science Network program of the Division of Materials Science and Engineering, 
US Department of Energy, grant no. DE-SC0007091.

\bibliographystyle{prsty}

\begin{thebibliography}{10}
\bibitem{notegap}
Let us note that these two different formulae for the orbiton spectral
  functions are not fully equivalent: Eqs.
  (\ref{eq:spectralbl_v2})-(\ref{eq:spectralcl_v2}) do not contain the orbital
  excitations which are above the Hubbard gap. However, we can safely neglect
  the coherent propagation of the orbital excitations in the Hilbert space {\it
  with} double occupancies, since: (i) for the coherent propagation of the
  orbital excitations the relevant gap is of the order of the Hubbard $U$ and
  not of the order of the charge transfer gap $\Delta$, and (ii)
  $\bar{\varepsilon}_b$ and $\bar{\varepsilon}_c$ are much small than the
  Hubbard $U$.
\bibitem{noteneelbethe}
If we were to make a similar analysis as above but for the spin sector of the
  ground state $|\Phi \rangle $ being a Neel AF instead of the quantum AF, we
  would notice that A and B2 processes are not present in that case and it is
  only the B1 process which contributes. Thus, we see that in that case the LOW
  approach works fine.
\end{thebibliography}

\appendix

\section{Reduction of the effective hopping in the linear chain due to $\propto t_n^2$ perturbative processes}
\label{sec:applambda}

Let us firstly state that, due to the $t_n^2$ processes, the number of holes residing
in the orbitals {\it within} the Cu-O-Cu-O-... chain for a particular CuO$_4$ cluster
depends on the particular $\alpha$ orbital forming the bonding state, cf. Fig.~\ref{fig:unitcell}.
More precisely, for the bonding states formed around the $b$ orbital
and occupied by one hole the whole charge is concentrated
in the orbitals within the Cu-O-Cu-O-... chain, while for the bonding states
formed by the $a$ or $c$ orbitals it is not the case. This is
because, in the latter case the bonding state is formed
by orbitals situated above and below the Cu-O-Cu-O-... chain. 

Looking in detail at this problem we concentrate first
at the case with the hole being initially doped into the $a$ orbital. 
The part of the charge transfer Hamiltonian~(\ref{eq:ct}) 
which is responsible for the effect mentioned above is:
\begin{align}
&- t_{\sigma o} \sum_{i, \sigma}\!\Big( f^\dag_{i a \sigma} 
f_{i yo+ \sigma}^{} -f^\dag_{i a \sigma}f_{i yo- \sigma}^{}  +\mbox{H.c.} \Big)  \nonumber \\
&+\Delta_{yo} \sum_{i} (n_{iyo+}+n_{iyo-}),
\end{align}
with $t_{\sigma o}$ being the main `actor' here, i.e. it is this 
hopping element which makes the hole escape from the Cu-O-Cu-O-... chain (see Fig. \ref{fig:unitcell}).
We can now easily diagonalize this three-level problem
and calculate the number of holes left in the $a$ orbital
by evaluating the following quantum mechanical amplitude
\begin{align}
 \lambda_a \equiv \langle a | \psi_a \rangle  = \frac{\Delta_{yo} - e_a}{\sqrt{2  t^2_{\sigma o} +(\Delta_{yo} - e_a)^2}},
\end{align}
where $e_a = (\Delta_{yo}-\sqrt{\Delta^2_{yo}+8t_{\sigma o}})/2$, while
$|\psi_a \rangle$ is the bonding state coming from
the above diagonalization procedure and which is well separated
from the antibonding and nonbonding states (so that
we can skip the latter two when studying the low energy regime). 

A similar analysis as above but for the $c$ orbital leads
to 
\begin{align}
 \lambda_c \equiv \langle c | \psi_c \rangle = \frac{\Delta_{xo} - e_c}{\sqrt{2  t^2_{\pi o} +(\Delta_{xo} - e_c)^2}},
\end{align}
where $e_c = (\Delta_{xo}-\sqrt{\Delta^2_{xo}+8t_{\pi o}})/2$.
Here again $|\psi_c \rangle$ is the bonding state
but this time centered around the $c$ orbital. 

Finally, since the $b$ orbital does not hybridize with the
oxygens lying above or below the Cu-O-Cu-O-... chain, the 
corresponding $\lambda_b$ would be equal
to unity and could be skipped in what follows.

These renormalized values of the number of holes within the chain
directly lead to renormalized values of the hopping elements from
the $\psi_a$ and $\psi_c$ orbitals with respect to the hopping
elements from the pure $a$ and $c$ orbitals. A rigorous
calculation would now require that together with using the parameters
$\lambda_a$ and $\lambda_c$ as renormalizing the hopping,
we should also use the basis spanned by the $\psi_a$ and 
$\psi_c$ orbitals. However, 
we avoid this in our calculations. We justify this `approximation' as follows:

In general, to properly account for {\it all} the effects arising
from the $\propto t_n^2$ perturbative processes, 
the rigorous treatment would require using the
so-called cell perturbation theory~\cite{Jefferson1992}: that is 
to rewrite the full charge transfer Hamiltonian using the bonding / antibonding
states and then to
calculate the superexchange interactions in this basis. This,
however, requires very tedious calculations as
even for the much simpler case of Ref. \onlinecite{Jefferson1992}
the problem is nontrivial and complex. 

Therefore, we follow the more standard route, i.e. we calculate
the superexchange interactions in Secs. \ref{sec:0b}-\ref{sec:0c} using the orbital basis
that was already used to write down the charge transfer model Eq.~(\ref{eq:ct}).
The only two remnants of the cell perturbation theory,
or in other words of the fact that the superexchange
should be modified due to the formation of bonding
and antibonding states, are: (i) the use of the
renormalized parameters $\bar{\varepsilon}_{\alpha}$ 
instead of ${\varepsilon}_{\alpha}$ as discussed in Sec.~\ref{sec:0a}, and (ii) the use of the factors
$\lambda_a$ and $\lambda_c$ which renormalize the 
number of holes present within the chain when
a hole is doped into $a$ or $c$ orbital, respectively (see above).
We have verified that the renormalization of other
parameters has a much smaller effect.
In particular: 
(i) the charge transfer energies in the bonding states should change
similarly for all orbitals with respect to their values
in the charge transfer model defined in the 2$p$ and 3$d$ orbital basis,
(ii) the matrix elements of the Coulomb interaction in the 
bonding / antibonding basis are similar to the ones calculated in the 2$p$
and 3$d$ orbital basis.

\section{Linear orbital wave approximation}
\label{sec:low}

The `standard' way to obtain the orbiton dispersion, in the spin-orbital model
is to use the linear orbital wave (LOW) approximation (cf. Ref. ~\onlinecite{Wohlfeld2009}) for the orbital pseudospin
degrees of freedom and to integrate out the spin degrees of freedom in a mean-field way~\cite{Wohlfeld2011}. 
This in general {\it may} be justified here due to the presence of the long range orbital order. 
In order to test this scenario, in this section of the appendix, we perform the LOW approximation and calculate
the orbiton spectral function (see part 1 below) together with the RIXS cross section (see part 2 below). 
We also discuss why the LOW approximation fails
in properly describing the experimental results reported in Ref.~\onlinecite{Schlappa2012}, cf. part 3 below.

\subsection{Spectral function in linear orbital wave approximation}

\label{sec:1b}
\begin{figure}[t!]
\centering
   \includegraphics[width=1.0\columnwidth]{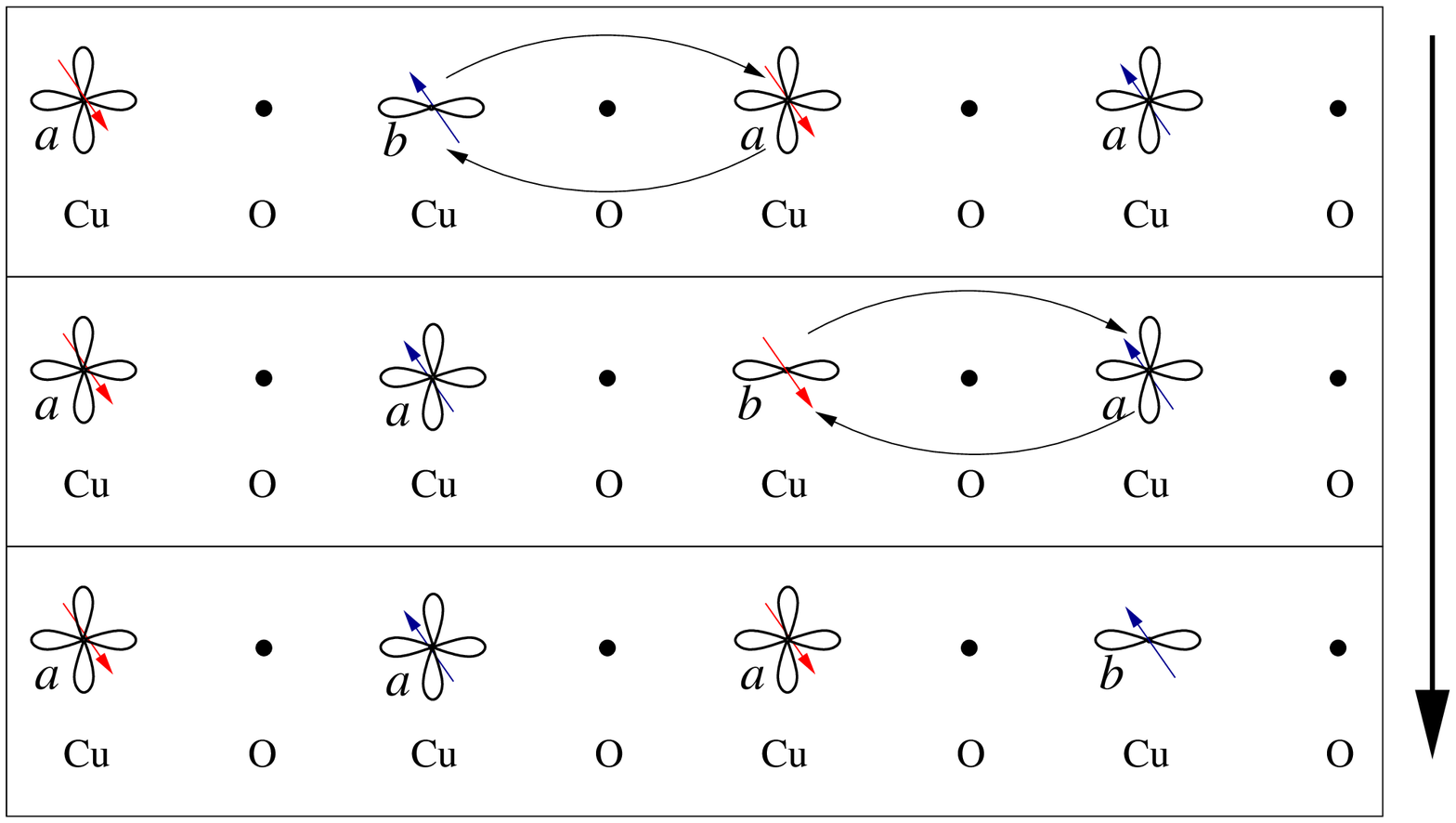}
\caption{(Color online) Schematic view of the propagation of
the orbiton in the LOW approximation: note that the orbiton moves in such a way that it does not disturb the AF correlations
which is due to the mean-field decoupling of spin and orbital degrees of freedom.   
}
\label{fig:propLOW}
\end{figure}

{\it LOW for $b$ orbiton.---} Following e.g. Ref.~\onlinecite{Wohlfeld2009} and Ref.~\onlinecite{Wohlfeld2011} 
we first introduce the following bosonic creation (annihilation) operators $\beta^\dag_j$ ($\beta_j$) for the orbital pseudospin operator:
\begin {align}\label{eq:sigmaboson}
\sigma^z_j & = \beta^\dag_j \beta_j - \frac12 , \nonumber \\
\sigma^+_j &= \beta^\dag_j, \nonumber \\
\sigma^-_j & = \beta_j,
\end{align}
where we already skipped the three-orbiton terms in the above expressions,
since we will keep only quadratic terms in the bosonic degrees of freedom in the effective Hamiltonian below. 
Besides, we decouple the orbital operators from the spins and assume for the spins
their appropriate mean field values. The latter is a standard procedure when calculating the spin wave
dispersion in the spin and orbitally ordered systems~\cite{Moussa1996, Oles2005}. Applying these transformations 
to the Hamiltonian $\mathcal{\bar{H}}$ we obtain the following LOW Hamiltonian:
\begin{align}
 H^{ab}_{\rm LOW} \equiv&  {H}^0_{\rm LOW} + {H}^a_{\rm LOW}+{H}^b_{\rm LOW} = 
 \sum_k (B+ 2 J_b \cos k)  \nonumber \\
& \times \beta^\dag_k \beta_k +J_1 (1+R) \sum_{i }  \left({\bf S}_{i } \cdot {\bf S}_{i+1} -
\frac{1}{4} \right), 
\end{align}
with the constants $B$ and $J_b$ defined as 
\begin{align} \label{eq:cm}
B \equiv & \bar{\varepsilon}_b -  \mathcal{A} (R^b_1 J^b_{12} + r^b_1 \frac{J_1+J^b_2}{2})  -  \mathcal{B} (R^b_2 J^b_{12} + r^b_2 \frac{J_1+J^b_2}{2}) \nonumber \\
&+ 2 J_1 (1+R)  \mathcal{B}
\end{align}
and
\begin{align}
J_b \equiv \frac12 J^c_{12} \big[\mathcal{A} (R^b_1 + r^b_1) -\mathcal{B} (R^b_2+r^b_2)\big],
\end{align}
with 
\begin{align}
 \mathcal{A} \equiv \Big \langle \Phi | {\bf S}_{i } \cdot {\bf S}_{i+1} +\frac{3}{4} | \Phi \Big\rangle,
\end{align}
and
\begin{align}
\mathcal{B}  \equiv \Big\langle \Phi | \frac{1}{4} - {\bf S}_{i } \cdot {\bf S}_{i+1} | \Phi \Big\rangle.
\end{align}
Here $|\Phi\rangle$ is the AF and FO ground state (cf. Sec. \ref{sec:1}) of $ H^b_{\rm LOW}$ with energy $E_\Phi$.
The values of the spin-spin correlations have to be calculated for the spin ground state of the Hamiltonian  $H^b_{\rm LOW}$
which is a quantum AF, see Sec. \ref{sec:1}. This can easily be obtained from the well-known exact Bethe-Ansatz-based solution for a 1D quantum AF: 
$\mathcal{A}= 0.31$ and $\mathcal{B}= 0.69$. (Let us note that these numbers are significantly 
different from the ones that are known for the not-realized-here `classical' case, i.e. for the ordered Neel AF -- in that case: $\mathcal{A}= 0.5$, $\mathcal{B}= 0.5$, and
$J_b \simeq 0.011$ eV.) Using the spin-orbital model parameters from Table~\ref{tab:1}, we finally obtain  $J_b \simeq -0.019$ eV as also reported
in Table~\ref{tab:1}.
%
\begin{figure}[t!]
   \includegraphics[width=0.4\textwidth]{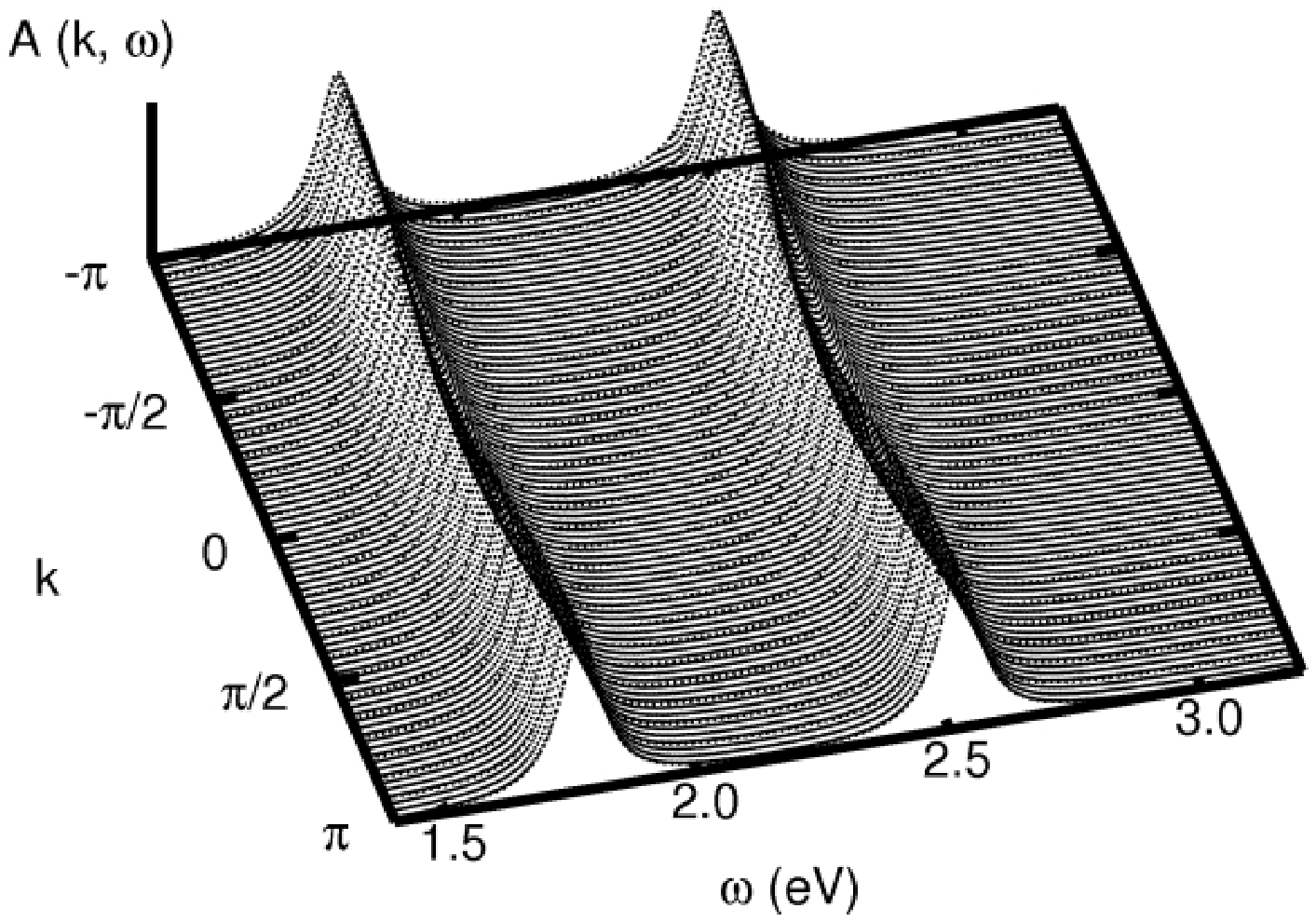}
   \includegraphics[width=0.4\textwidth]{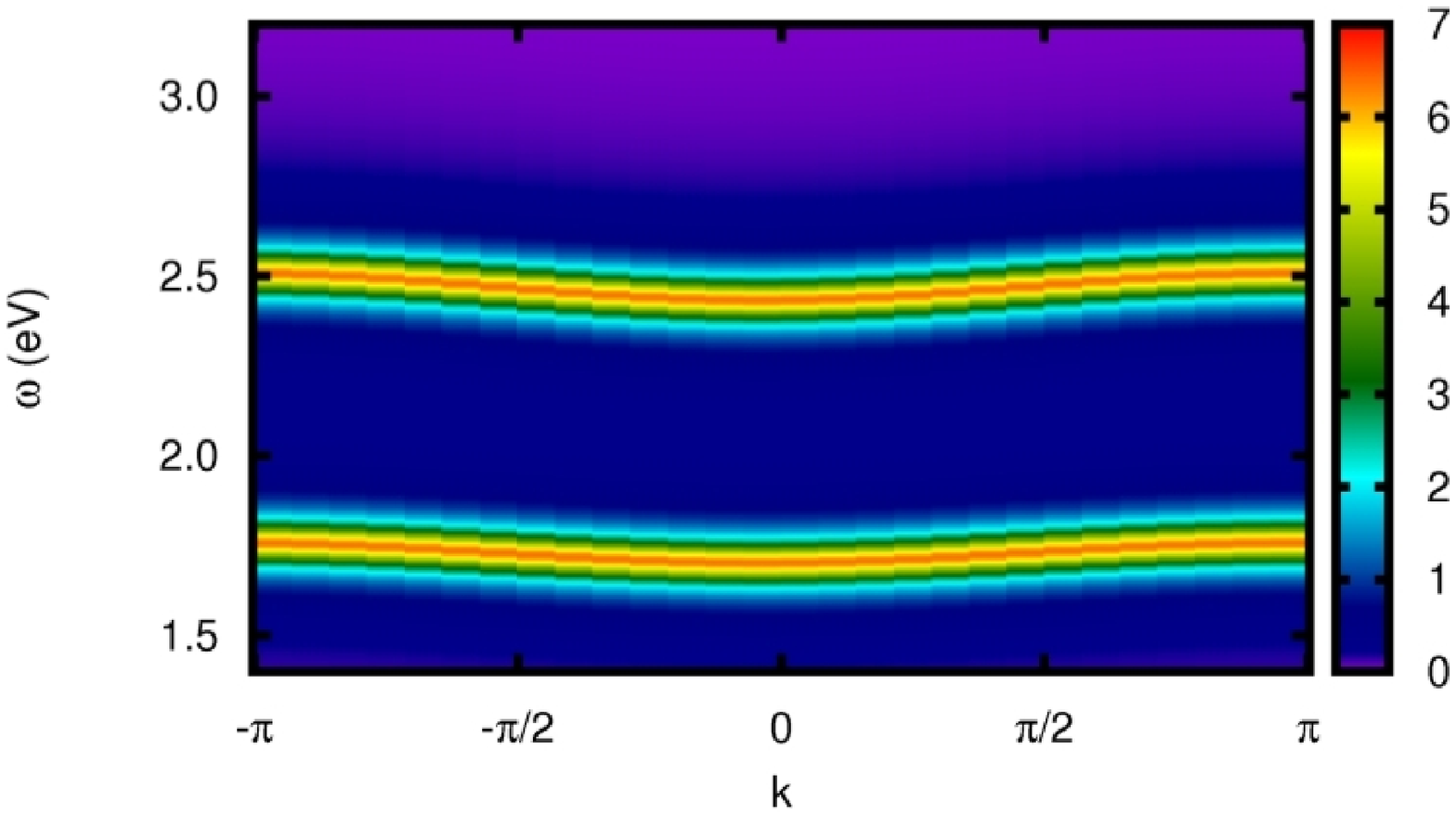}
\caption{(Color online) Spectral function $A_b(k, \omega) + A_c(k, \omega)$ as a function of momentum $k$ and  
energy transfer $\omega$ in the LOW approximation and quantum AF case. 
Results for broadening $\eta =0.05$ eV (cf. caption of Fig. \ref{fig:spectrumSOS}).
}
\label{fig:spectrumLOW}
\end{figure}

Although one could directly use Eq. (\ref{eq:cm}) to calculate the on-site cost of an orbital excitation $B$, this value can also be calculated by 
using the {\it ab-initio} quantum chemistry calculation for a {\it ferromagnetic} chain with four CuO$_3$
plaquettes. The latter gives the value of a single orbital excitation in the ferromagnetic chain $E_b=2.15$ eV (cf. discussion in Sec. \ref{sec:1} 
and Table \ref{tab:1}) and leads to 
\begin{align} \label{eq:ec}
B \simeq E_b + E_{AF} 
\end{align}
and, since $E_{AF}= 0.33$ eV,  $B\simeq 2.48$ eV, cf. Table \ref{tab:1}. To be in line with Ref.~\onlinecite{Schlappa2012}, we use the latter
value as the cost of the local $b$ orbital excitations in the AF chain. 

Having defined the Hamiltonian and its parameters, we are now ready to compute the orbiton spectral function 
$A_b(k, \omega)$  [Eq. (\ref{eq:spectralbl_v2})], which, when expressed in the new bosonic operators [Eq. ($\ref{eq:sigmaboson}$)],
reads
\begin{align}
A_b(k, \omega)=\frac{1}{\pi} \lim_{\eta \rightarrow 0} 
\Im \langle \Phi | \beta_k \frac{1}{\omega + E_\Phi - H^{ab}_{\rm LOW} - i \eta } \beta_k^\dag | \Phi \rangle.
\end{align}
Since $|\Phi\rangle$ is a vacuum for boson operators $\beta |\Phi\rangle =0$, we easily 
obtain
\begin{align}
A_b(k, \omega)=\frac{1}{\pi}  \lim_{\eta \rightarrow 0} 
\Im \frac{1}{\omega - B-2J_b\cos k - i \eta }.
\end{align}
The orbiton spectral function consists
of a single quasiparticle peak with a sine-like dispersion, with period $2 \pi $ and bandwidth $4 |J_b| \simeq 0.08$ eV, 
cf. Fig.~\ref{fig:spectrumLOW}. 
This result can be intuitively understood by looking at the cartoon picture of the orbiton propagation
in the LOW approximation, cf. Fig. \ref{fig:propLOW}.

{\it LOW for $c$ orbiton.---} Following the same steps as above we obtain:
\begin{align}
 H^{ac}_{\rm LOW} \equiv&  {H}^0_{\rm LOW} + {H}^a_{\rm LOW}+{H}^c_{\rm LOW} = 
 \sum_k (C+ 2 J_c \cos k)  \nonumber \\
& \times  \beta^\dag_k \beta_k
+J_1 (1+R) \sum_{i }  \left({\bf S}_{i } \cdot {\bf S}_{i+1} -
\frac{1}{4} \right)
\end{align}
with the constants $C$ and $J_c$ defined as 
\begin{align}
C \equiv & \bar{\varepsilon}_c -  \mathcal{A} (R^c_1 J^c_{12} + r^c_1 \frac{J_1+J^c_2}{2})  -  \mathcal{B} (R^c_2 J^c_{12} + r^c_2 \frac{J_1+J^c_2}{2}) \nonumber \\
&+ 2 J_1 (1+R)  \mathcal{B}
\end{align}
and
\begin{align}\label{eq:jc}
J_c \equiv \frac12 J^c_{12} \big[\mathcal{A} (R^c_1 + r^c_1) -\mathcal{B} (R^c_2+r^c_2)\big].
\end{align}
which gives $J_c \simeq -0.014$ eV, cf. Table~\ref{tab:1}.
Again $C\simeq 1.74$ eV can be estimated using the {\it ab-initio} calculated value (cf. discussion above for the $b$ orbiton) of $E_c \simeq 1.41$ eV
for a local $c$ orbital excitation in the ferromagnetic chain
and the relation (see above)
\begin{align}
C \simeq E_c + E_{AF}.
\end{align}
Finally, we obtain the following spectral function $A_c(k, \omega)$
\begin{align}
A_c(k, \omega)=\frac{1}{\pi}  \lim_{\eta \rightarrow 0} 
\Im \frac{1}{\omega - C-2J_c\cos k - i \eta },
\end{align}
which is also shown in Fig.~\ref{fig:spectrumLOW} and which qualitatively resembles
the above calculated $b$ orbiton dispersion.

{\it Spin-orbital waves.---} As a side remark, let us note that the joint spin-orbital wave defined as in
Ref. \onlinecite{Oles2000} cannot be present in the considered spin-orbital model. 
The reason is that this would require such terms as e.g. $(S^+_i S^-_j + S^-_i S^+_j)  (\tau^+_i \tau^+_j + \tau^-_i \tau^-_j) $ 
to be present in the spin-orbital Hamiltonian (\ref{eq:hspinorb}) -- which is not the case here.

\subsection{RIXS in linear orbital wave scenario}
\begin{figure*}[t!]
   \includegraphics[width=0.4\textwidth]{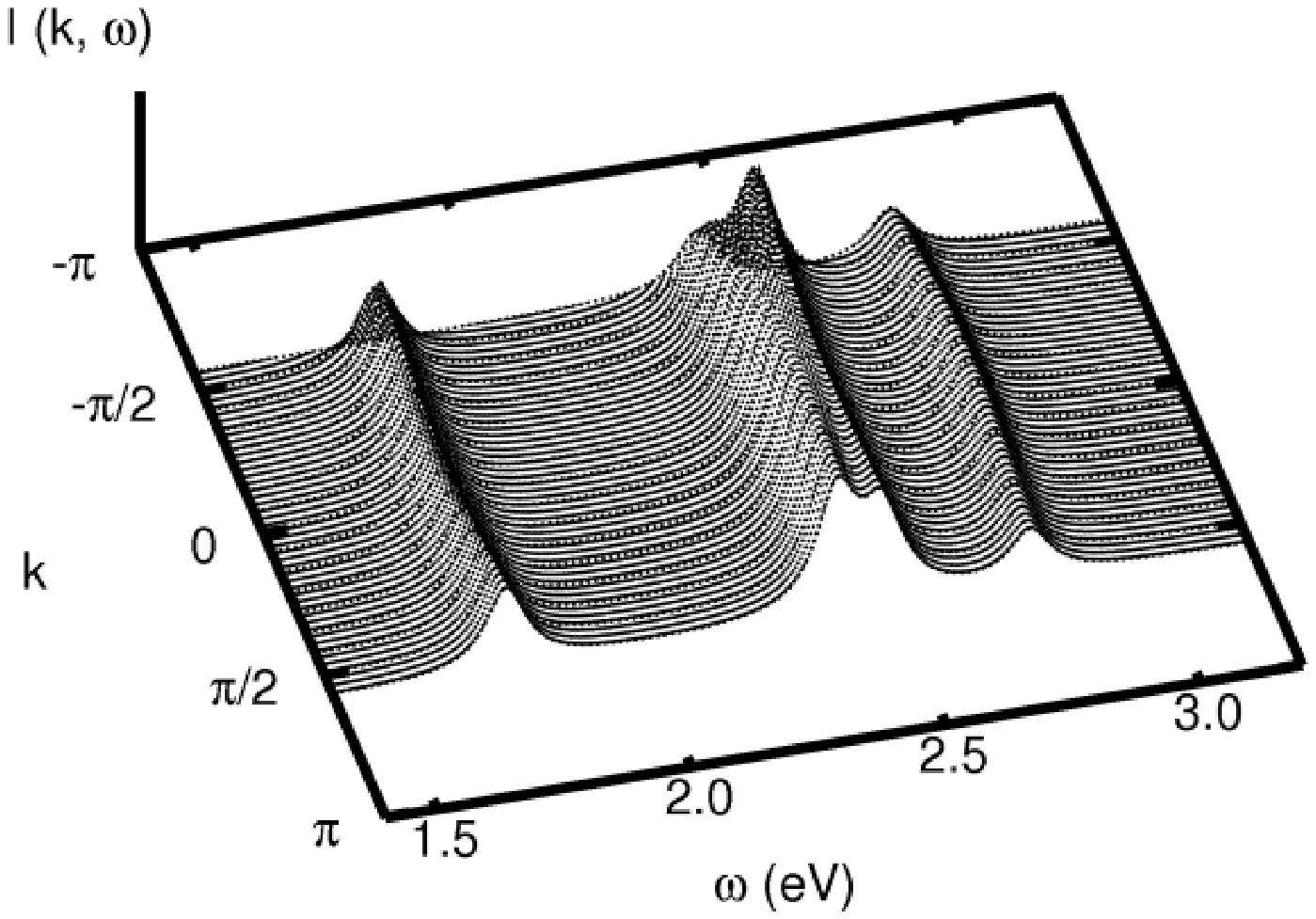}
   \includegraphics[width=0.4\textwidth]{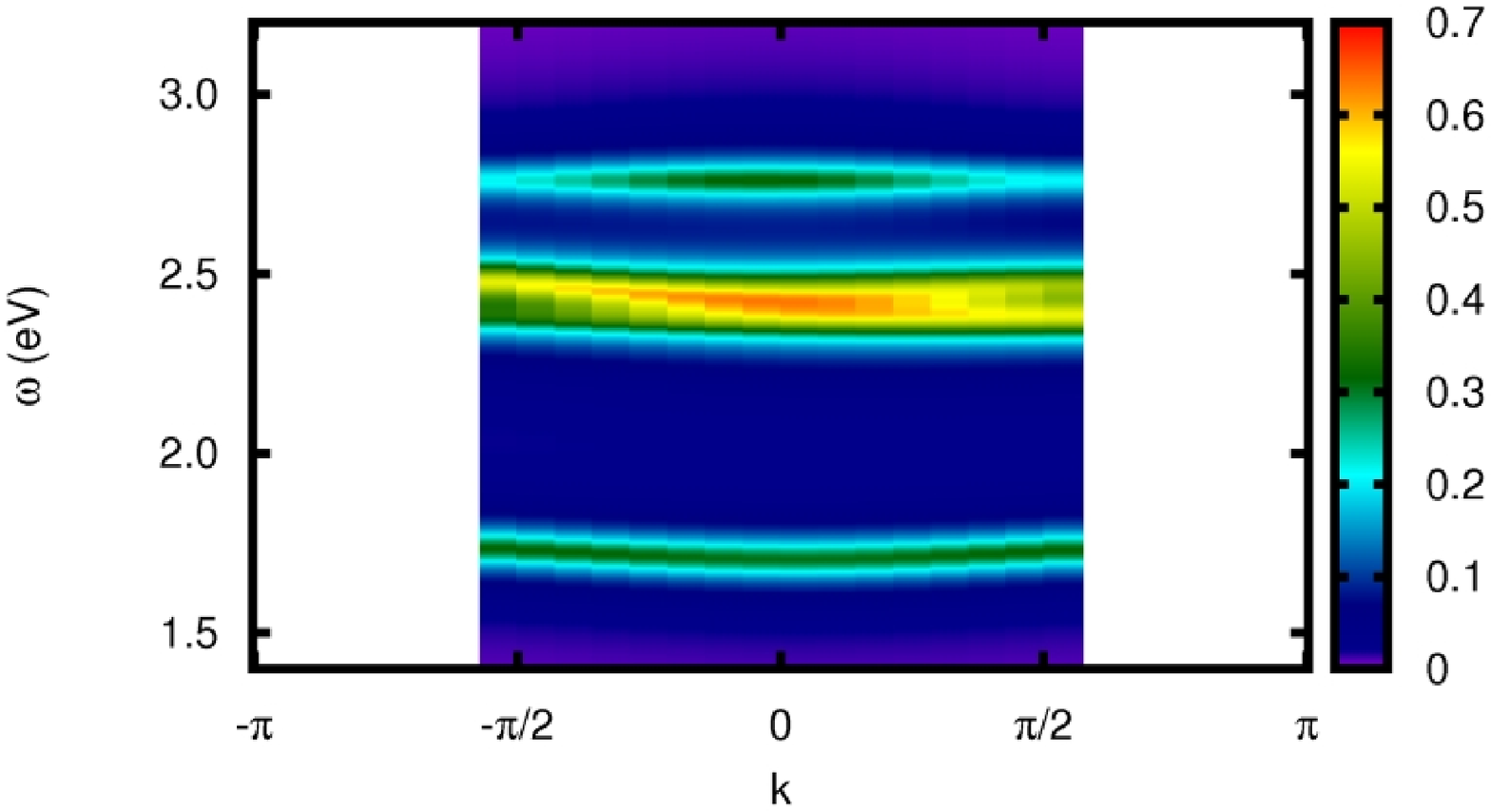}
\\[1.0em]
   \includegraphics[width=0.4\textwidth]{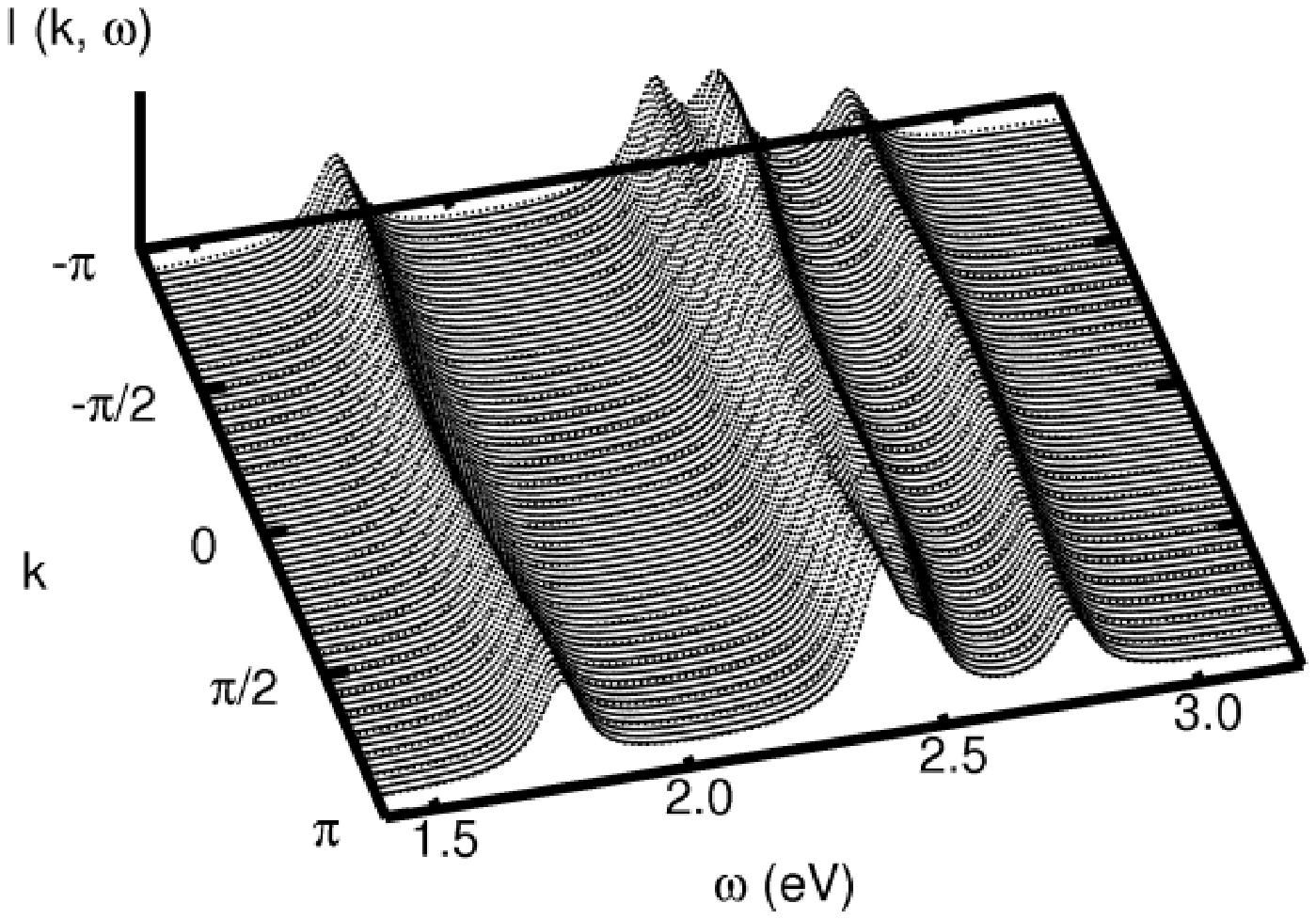}
   \includegraphics[width=0.4\textwidth]{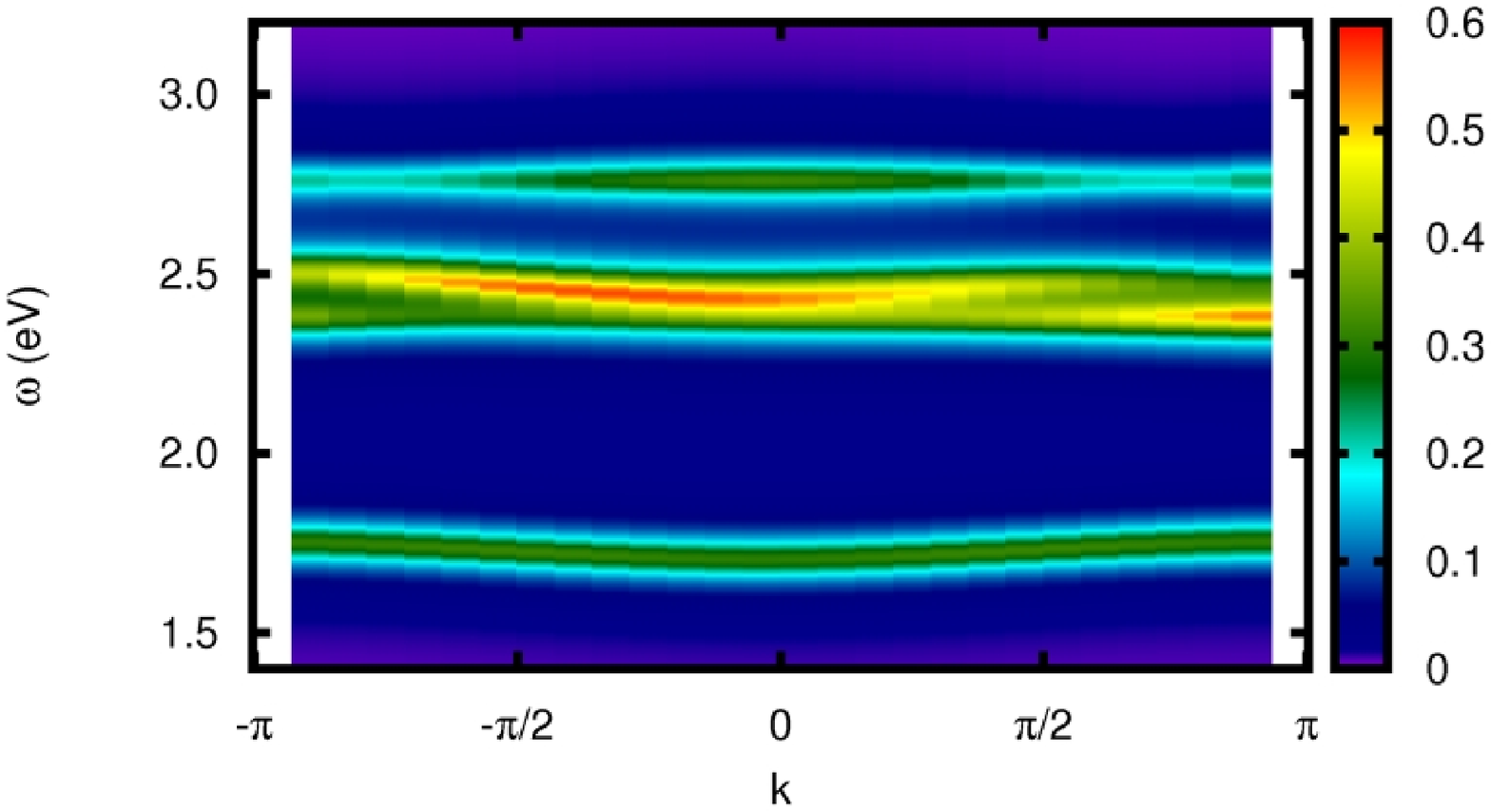}
\caption{(Color online) RIXS cross section for 90$^0$ (130$^0$) scattering 
geometry as calculated in the LOW approximation and convoluted
with the results from the local model, Fig. \ref{fig:localRIXS}, 
on the top (bottom) panel.
Left (right) panels show line (color map) spectra.
Results for broadening $\eta =0.05$ eV (cf.
caption of Fig. \ref{fig:spectrumSOS}).
}
\label{fig:LOWRIXS}
\end{figure*}
In order to calculate RIXS cross section in the LOW approximation we first express
the RIXS operator Eq.~(\ref{eq:tc}) in terms of the orbital pseudospins. For the $b$
orbiton case (i.e. $T_b$ operator), following Eq. (\ref{eq:pseudospinb}) and Eq. (\ref{eq:spin}), we obtain
\begin{align}\label{eq:tbl}
T_b =&  \frac{1}{\sqrt{N}} \sum_{j} e^{ikj} [ (B_{\uparrow, \uparrow}+B_{\downarrow, \downarrow}) \frac12 \sigma^+_j +  
(B_{\uparrow, \uparrow}-B_{\downarrow, \downarrow}) S^z_j \sigma^+_j \nonumber \\
 &+ B_{\uparrow, \downarrow}S^+_j \sigma^+_j + B_{\downarrow, \uparrow} S^-_j \sigma^+_j]
\end{align}
Next, following Eq. (\ref{eq:sigmaboson}), we express the {\it first term} of the above written 
RIXS operator in terms of the Holstein-Primakoff bosons $\beta^+_k$
\begin{align}\label{eq:tcc}
 T^{(1)}_b =  \frac12 (B_{\uparrow, \uparrow}+B_{\downarrow, \downarrow}) \beta^+_k .
\end{align}
Note that in the above expression the spin-dependent part [the three last terms of the right hand side of Eq. (\ref{eq:tbl})] 
in Eq. (\ref{eq:tcc}) is skipped. This is because it does not lead to any
dispersive excitations since there are no terms in the LOW Hamiltonian which could
move a spin excitation together with an orbital excitation (in the LOW approximation). 
(This is somewhat similar to the problem of a hole doped into the orbitally ordered state
in a 1D chain which can `visit' the neighbouring sites but which spectrum is 
$k$-independent~\cite{Wohlfeld2008}.) However, these three neglected terms will 
contribute to the total RIXS cross section as dispersionless excitations. 

When, apart from the above discussed $b$ orbiton case, we also include the contribution from
the $c$ orbiton (which is analogous to the $b$ orbiton case) and from the dispersionless
$d$ and $e$ orbitons, we obtain
\begin{align} \label{eq:rixslow}
I(k, \omega)=& \frac14 |B_{\uparrow, \uparrow}+B_{\downarrow, \downarrow}|^2 A_b (k, \omega) \nonumber \\
&+ \Big(\frac14 |B_{\uparrow, \uparrow}-B_{\downarrow, \downarrow}|^2+ |B_{\downarrow, \uparrow} |^2 \Big) \delta(\omega - E_b- E_{AF}) \nonumber \\
&+\frac14 |C_{\uparrow, \uparrow}+C_{\downarrow, \downarrow}|^2 A_c (k, \omega) \nonumber \\
&+ \Big(\frac14 |C_{\uparrow, \uparrow}-C_{\downarrow, \downarrow}|^2+ |C_{\downarrow, \uparrow} |^2 \Big) \delta(\omega - E_c- E_{AF}) \nonumber \\
&+(|D_{\uparrow, \uparrow}({ k})|^2+|D_{\uparrow, \downarrow}({ k})|^2) \delta(\omega - E_d- E_{AF}) \nonumber \\ 
&+(|E_{\uparrow, \uparrow}({ k})|^2+|E_{\uparrow, \downarrow}({ k})|^2) \delta(\omega - E_e- E_{AF}).
\end{align}
Here the spectral functions $A_b (k, \omega)$ and $ A_c (k, \omega)$ are calculated in part 1 of Appendix~\ref{sec:low} and
the RIXS matrix elements follow from Sec.~\ref{sec:matrix}.

{\it Comparison with the experiment.---}
The RIXS cross section calculated using Eq. (\ref{eq:rixslow})
is shown in Fig. \ref{fig:LOWRIXS}. A small dispersion of the orbiton
excitations, known already from the spectral functions in part 1 of Appendix~\ref{sec:low}, is relatively well visible in the RIXS
cross section. However, there is a {\it qualitative} disagreement between these theoretical
calculations and the experimental results, cf. Fig. 4(a) in Ref.~\onlinecite{Schlappa2012}
for the case of the RIXS with scattering angle $\Psi = 130^0$; a similar disagreement is obtained
for the unpublished RIXS experimental results~\cite{privcomm} for the scattering angle $\Psi = 90^0$.
The main differences are as follows:
(i) the theoretical dispersion has its minimum at $k=0$ according to the calculations
while this is not the case in the experiment,
(ii) the theoretical
results do not predict the onset of a continuum above the $b$
excitation,
(iii) the obtained dispersion of the quasiparticle peaks
is much smaller than in the experiment,
and (iv) there is a disagreement 
between the calculated and measured RIXS intensities.

\subsection{Why linear orbital wave approximation fails}
\begin{figure}[t!]
\includegraphics[width=1.0\columnwidth]{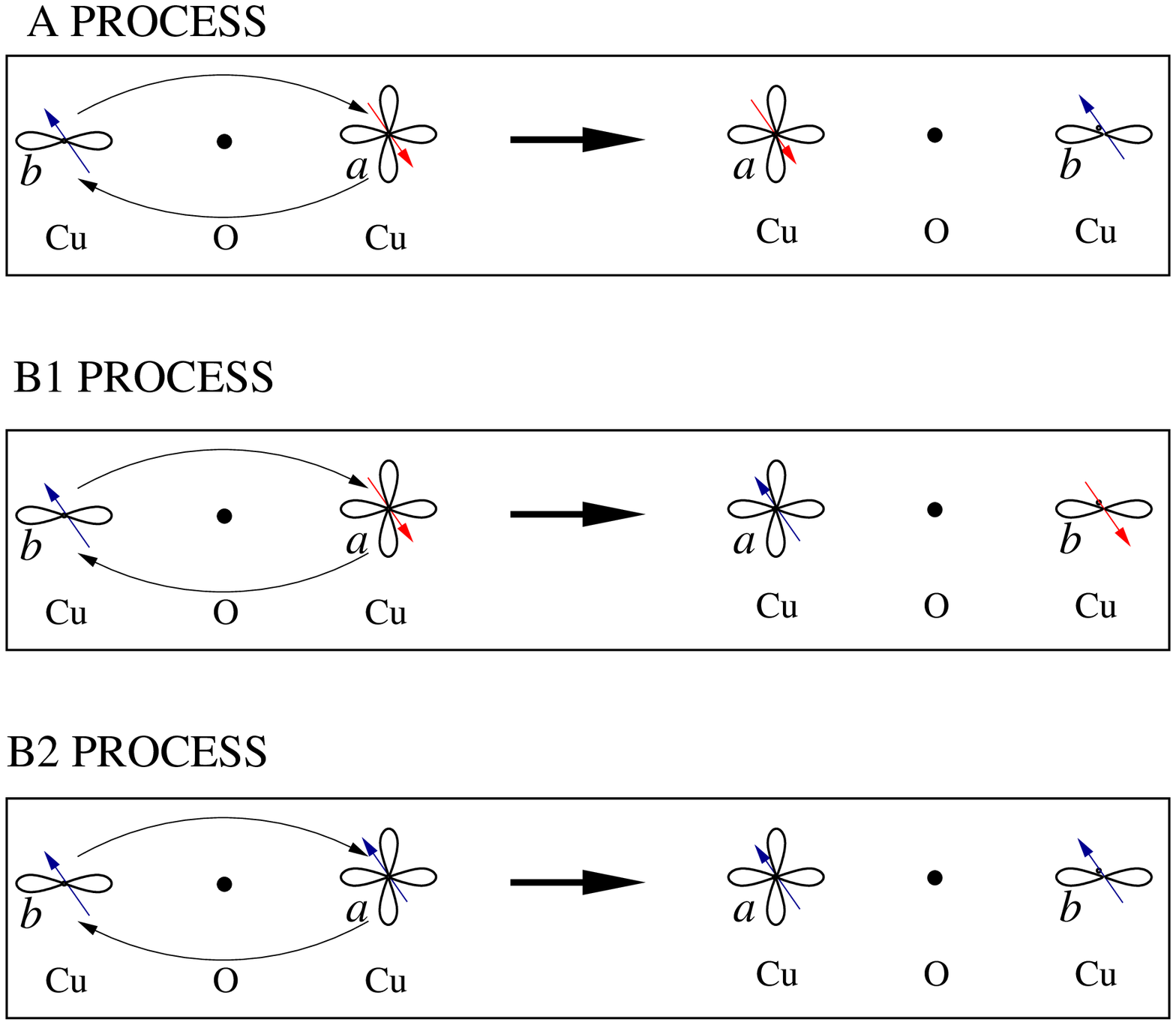}
\caption{(Color online) Schematic view of the three possible superexchange 
processes which lead to the orbiton propagation (here shown
for the $b$ orbiton but the $c$ orbiton case is analogous). 
In the LOW approximation all of these processes, which amplitudes
have different signs, are summed up.
When treated separately, as in the spin-orbital separation 
approach, it occurs that for the orbiton propagation 
in a quantum AF only processes A and B2 matter.
}
\label{fig:a_b1_b2_processes}
\end{figure}
{\it General considerations.---}
Let us show why the LOW theory, employed above in calculating the orbiton spectral
function, cannot correctly reproduce the orbiton propagation in the here discussed spin-orbital model.

In order to do that, we take a closer look at all possible channels of the orbiton propagation,
see Fig. \ref{fig:a_b1_b2_processes}, and calculate their
relative contribution to the orbiton propagation in the LOW approximation. (In what follows
we concentrate on the $b$ orbiton case but similar arguments apply to the $c$ orbiton
case.) Thus, we split the effective orbiton superexchange process $\propto J_b$, as calculated
in the previous section Eq. (\ref{eq:jc}), into three different contributions:
\begin{align}
 J_b = t_A + t_{B1} + t_{B2},
\end{align}
where
\begin{align}
 t_A = \frac12 J^b_{12} (R^b_1 +r^b_1+ R^b_2 + r^b_2)   \Big\langle \Phi \Big| \frac12 ({S}^+_{i } {S}^-_{i+1} + h.c.) \Big| \Phi \Big\rangle,
\end{align}
and
\begin{align}
 t_B =& t_{B1} + t_{B2}= \frac12 J^c_{12} \Big[ (R^b_1 +r^b_1)   \Big\langle \Phi \Big| {S}^z_{i } {S}^z_{i+1} +\frac34 \Big| \Phi \Big\rangle \nonumber \\
&- (R^b_2 + r^b_2) \Big\langle \Phi \Big| \frac14 -{S}^z_{i } {S}^z_{i+1}  \Big| \Phi \Big\rangle  \Big],
\end{align}
with
\begin{align}
 t_{B1} =& \frac12 J^b_{12} \mu \Big[ (R^b_1 +r^b_1)  \Big\langle \uparrow \downarrow \Big| {S}^z_{i } {S}^z_{i+1} +\frac34 \Big| \uparrow \downarrow  \Big\rangle \nonumber \\
&- (R^b_2 +r^b_2)  \Big\langle \uparrow \downarrow \Big| \frac14 - {S}^z_{i } {S}^z_{i+1}  \Big| \uparrow \downarrow  \Big\rangle \Big],
\end{align}
and
\begin{align}
 t_{B2} = \frac12 J^b_{12} (R^b_1 +r^b_1)  \nu \Big\langle \downarrow \downarrow \Big| {S}^z_{i } {S}^z_{i+1} +\frac34 \Big| \downarrow \downarrow  \Big\rangle, 
\end{align}
where $|\downarrow \downarrow \rangle$ denotes a ferromagnetic state,
 $|\downarrow \uparrow \rangle$ denotes a Neel AF state,
 and $\mu = |\langle \Phi | \downarrow \uparrow \rangle |^2 \sim 0.8$  and $\nu = |\langle \Phi | \downarrow \downarrow \rangle |^2 \sim 0.2$.
Substituting parameters from Table \ref{tab:1} and spin correlations for the quantum AF, Neel 
AF, and ferromagnetic state we obtain that
$t_A \sim -0.046$ eV, $t_{B1} \sim 0.009$ eV, and $t_{B2} \sim 0.018$ eV
(one can check that altogether they indeed give $J_b \sim -0.019$ eV as calculated in the previous section, cf. Table \ref{tab:1}).
Thus, we see that: (i) the A process has a surprisingly large
contribution and an opposite sign to the other processes -- so, unlike in the LOW result
presented above, we should treat it separately as we make a huge error when we
add all of these processes together, (ii) the B1 process is not only much smaller 
than the A process but also it is twice smaller than the B2 process~\cite{noteneelbethe}. 

This means that it is reasonable to try to define such an approximation, when calculating the orbiton propagation in 
the spin-orbital model (\ref{eq:hspinorb}), that, unlike the LOW approximation, will not average over these three processes. At the same time, such approximation
could neglect the B1 process due to its relatively small amplitude. 
In order to verify what kind of approximation can be used, let us try
to intuitively understand the difference between these three processes, cf. Fig. \ref{fig:a_b1_b2_processes}.
While process A denotes an orbiton hopping accompanied by a spin flip, the B processes
describe orbiton hoppings without any change in the spin background: the B1 for the case
when the spins on the bond are antiparallel, while the B2 when the spins are parallel. 
However, one can also look at this problem in a different way:
for the A and B2 process the spin of the hole in orbital $b$ is conserved during the spin-orbital exchange.
Hence, when process B1 is neglected, one can safely assume that the spin of the hole in the excited orbital does not
change during orbiton propagation -- and this is  the essence of the mapping to the $t$--$J$ model discussed
in the main text of the paper.

\section{RIXS for dispersionless orbital excitations}
\begin{figure*}[t!]
\begin{center}
\includegraphics[width=0.45\textwidth]{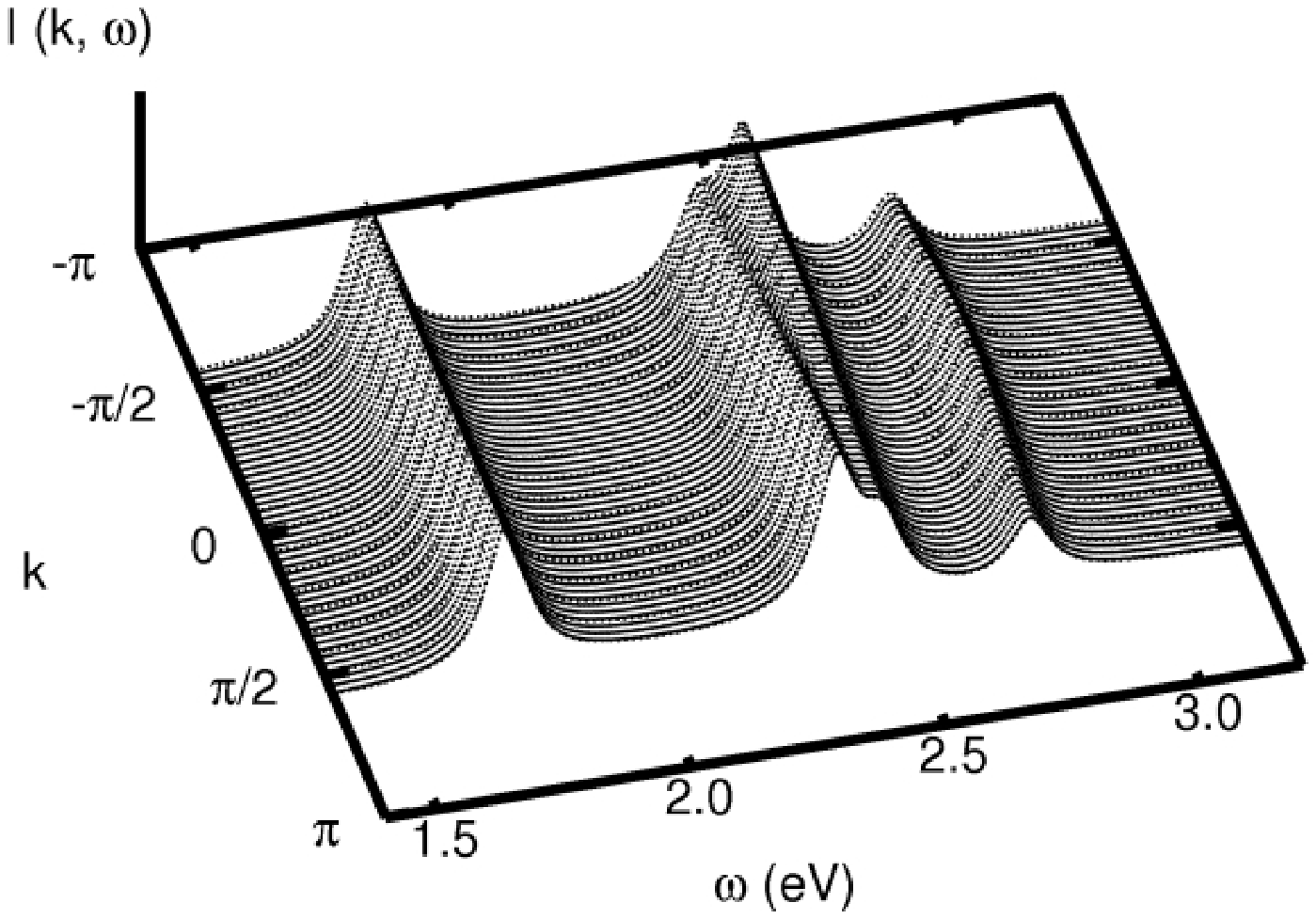} 
\includegraphics[width=0.45\textwidth]{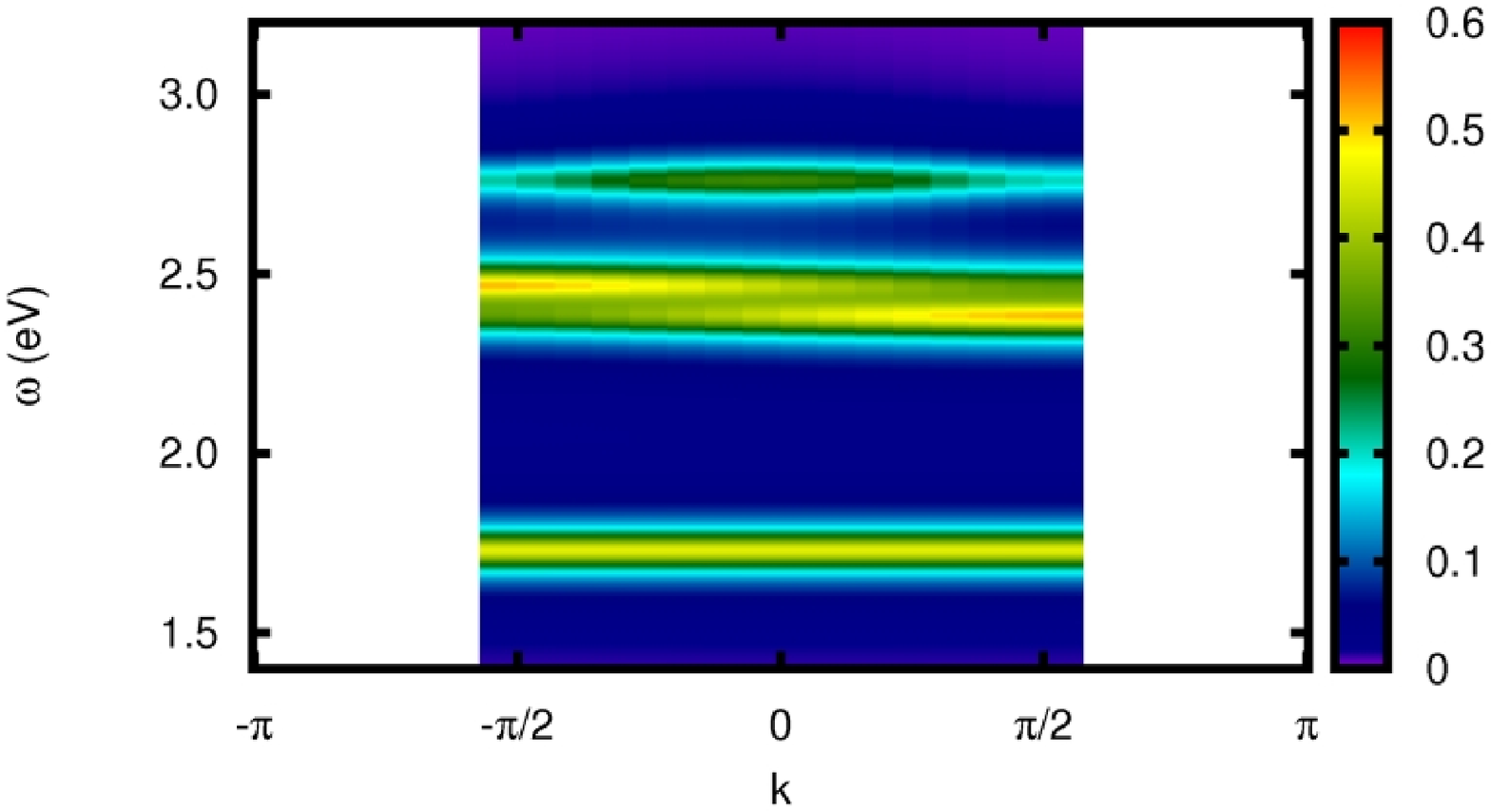} \\
\includegraphics[width=0.45\textwidth]{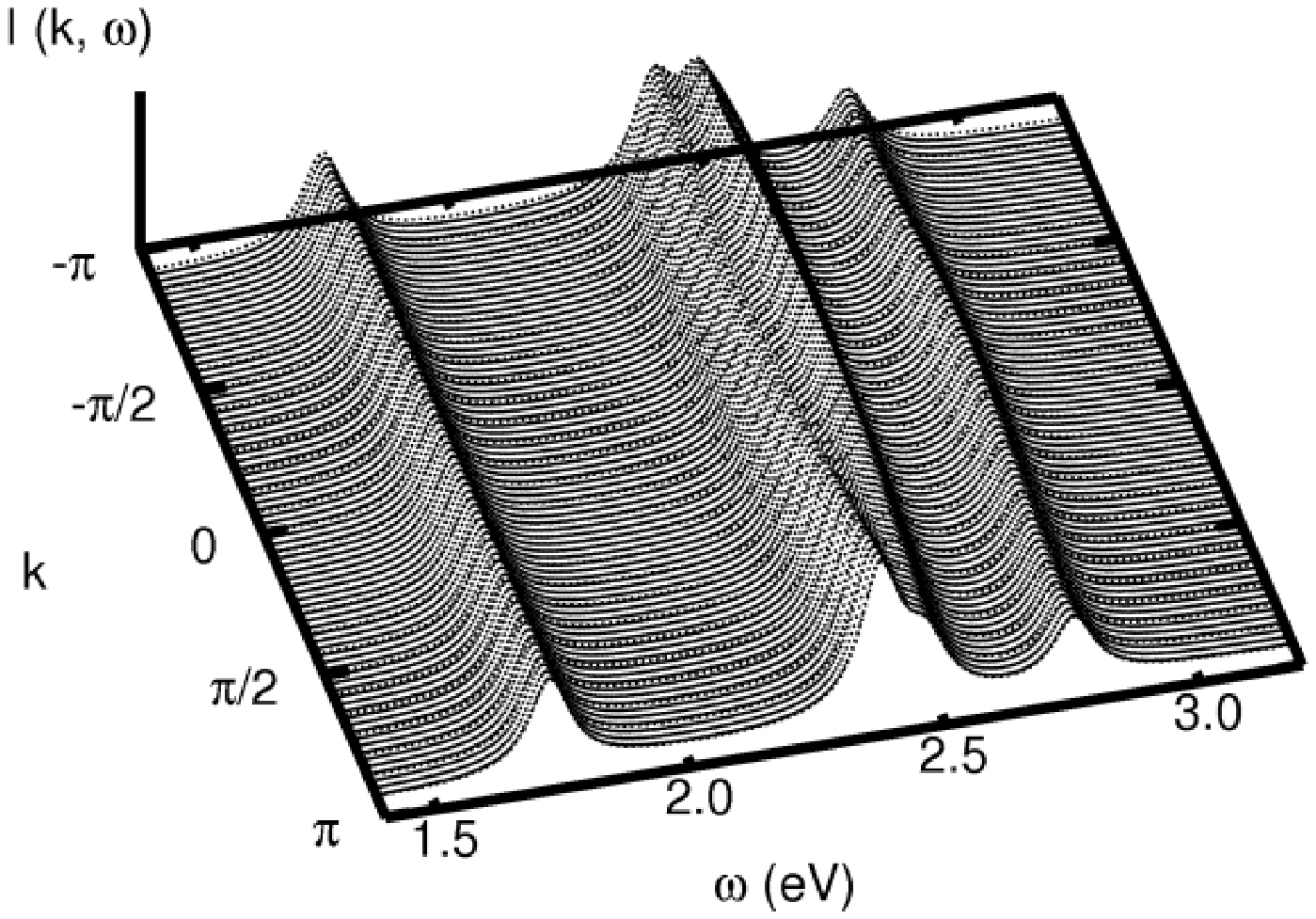} 
\includegraphics[width=0.45\textwidth]{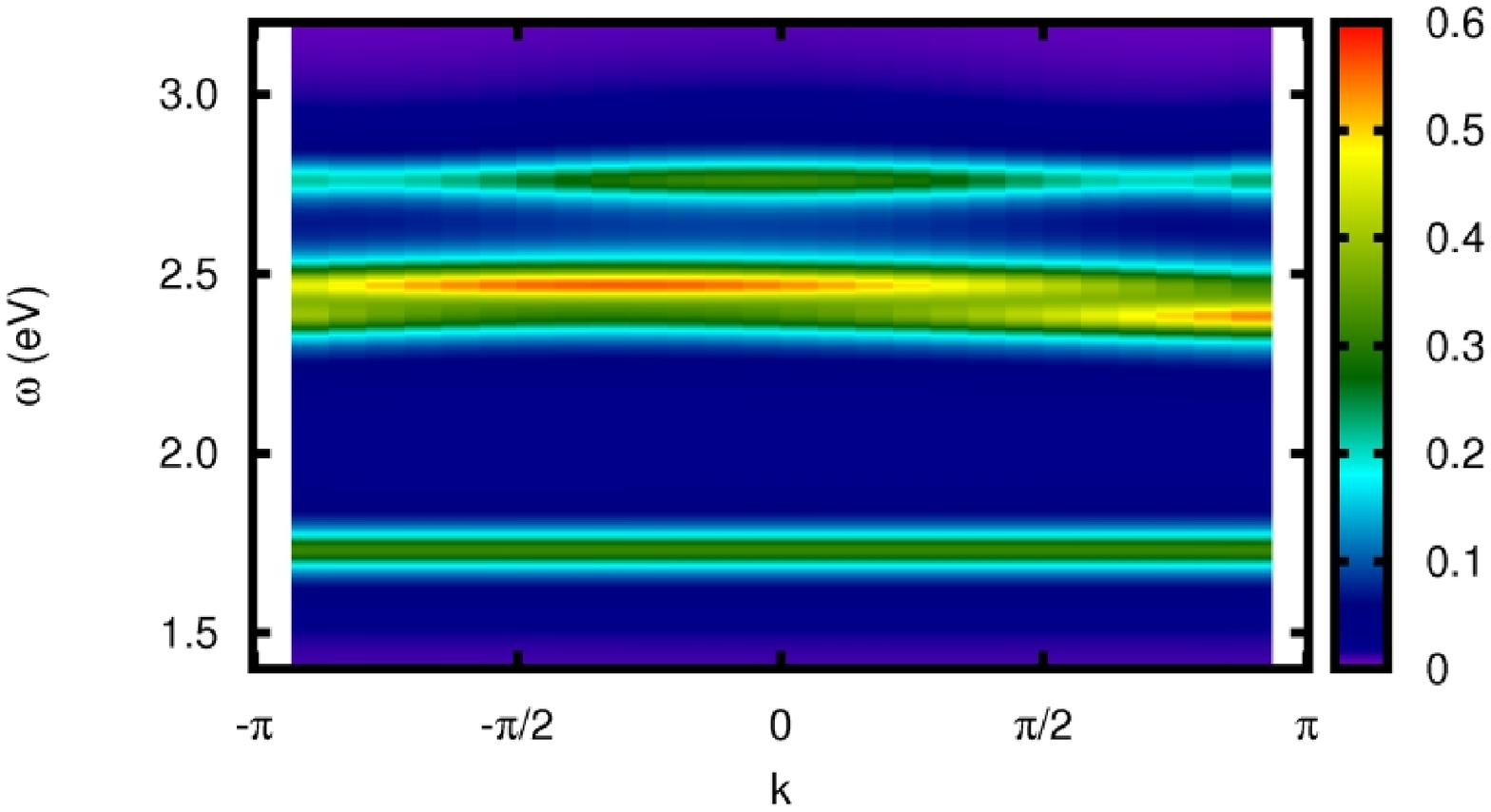} 
\end{center} 
\caption{(Color online) RIXS cross section for 90$^0$ (130$^0$) scattering 
geometry as calculated in the local model on the top (bottom) panel.
Left (right) panels show line (colour map) spectra. Results for broadening
$\eta =0.05 $ eV (cf. caption of Fig. \ref{fig:spectrumSOS}).
}
\label{fig:localRIXS}
\end{figure*}
\label{sec:2a}
In order to verify that  the RIXS cross section in the so-called `local model', i.e. with {\it all} orbital excitations
dispersionless, indeed does not follow the experimental RIXS
cross section~\cite{Schlappa2012}, we study the RIXS response for the following `local'
Hamiltonian
\begin{align} \label{eq:local}
 \mathcal{{H}} = &(E_b+ E_{AF})\sum_{i} ( n_{i b } -  n_{i a } ) \nonumber \\
 &+ (E_c +  E_{AF})\sum_{i} (n_{i c } -  n_{i a } ) \nonumber \\ &+ (E_d + E_{AF} ) \sum_{i} ( n_{i d }-  n_{i a } ) \nonumber \\
& + (E_e + E_{AF} ) \sum_{i} ( n_{i e }-  n_{i a } ).
\end{align}
Here $E_b$, $E_c$, $E_d$, and $E_e$ are the costs of the local orbital excitations as calculated using the {\it ab-initio} quantum chemistry
cluster calculations for the {\it ferromagnetic} CuO$_3$ chain in Sr$_2$CuO$_3$ 
(see Table \ref{tab:1}), while $E_{AF} = 0.24$ eV is the estimated correction to these values due to the quantum
AF ground state (see Table \ref{tab:1}). Substituting Eq. (\ref{eq:local}) into Eq. (\ref{eq:i}) and using Eq. (\ref{eq:tc}) 
we easily obtain
\begin{align} \label{eq:intensity_local}
 I ({ k}, \omega) &= 
(|B_{\uparrow, \uparrow}({ k})|^2+|B_{\uparrow, \downarrow}({ k})|^2) \delta(\omega - E_b- E_{AF}) \nonumber \\
&+ (|C_{\uparrow, \uparrow}({ k})|^2+|C_{\uparrow, \downarrow}({ k})|^2) \delta(\omega - E_c- E_{AF}) \nonumber \\
&+(|D_{\uparrow, \uparrow}({ k})|^2+|D_{\uparrow, \downarrow}({ k})|^2) \delta(\omega - E_d- E_{AF}) \nonumber \\ &+(|E_{\uparrow, \uparrow}({ k})|^2+|E_{\uparrow, \downarrow}({ k})|^2) \delta(\omega - E_e- E_{AF}),
\end{align}
which is shown in Fig. \ref{fig:localRIXS} for the two discussed scattering geometries.
It can be easily verified that the cross section calculated in this way does not agree with the
RIXS experimental cross section, as shown in Fig. 4(a) in Ref. \onlinecite{Schlappa2012}
(for the case of the scattering angle $\Psi = 130^0$; a similar disagreement is obtained
for the unpublished RIXS experimental results~\cite{privcomm} for the scattering angle $\Psi = 90^0$).

\end{document}